%% file: CaptchasITW.tex
\documentclass[letterpaper,twocolumn,10pt]{article}
\usepackage{usenix-2020-09}

\usepackage{tikz}
\usepackage{amsmath}
\usepackage{paralist}
\usepackage{enumitem}

\usepackage[centerlast]{caption}
\usepackage[section]{placeins}
\usepackage{tikz}
\usepackage{array}
\usepackage{booktabs}
\usepackage{longtable}
\usepackage{multirow}
\usepackage{stfloats}
\usepackage{subcaption}
\usepackage{tabularx}
\usepackage{tabulary}
\usepackage{xurl}
\usepackage{hyperref}
\usepackage{float}
\usepackage{xspace}

\usepackage{enumitem}
\setlist[itemize]{leftmargin=*}
\setlist[enumerate]{leftmargin=*}

\newcommand{\Captcha}{\textsc{Captcha}\xspace}
\newcommand{\Captchas}{\textsc{Captchas}\xspace}
\newcommand{\captcha}{\textsc{captcha}\xspace}
\newcommand{\captchas}{\textsc{captchas}\xspace}

\newcommand{\taggedpara}[1]{\textbf{#1}\xspace}

\ifdefined\showcomments
\usepackage{todonotes}
\usepackage{letltxmacro}
\LetLtxMacro{\todonote}{\todo}
\renewcommand{\todo}[2][]
{\todonote[inline, caption={#2}, size=\footnotesize, #1]
	{\renewcommand{\baselinestretch}{0.5}\selectfont#2\par}}
\else
\usepackage[disable]{todonotes}
\fi

\ifdefined\showchanges

\else

\fi

\begin{document}	
    	
	\title{An Empirical Study \& Evaluation of Modern CAPTCHAs}

	\newcolumntype{Y}{>{\centering\arraybackslash}X}

\author{
\begin{tabularx}{\textwidth}{ Y Y Y }
Andrew Searles                   & Yoshimichi Nakatsuka\footnotemark{}            & Ercan Ozturk \\
UC Irvine & ETH Z\"urich & UC Irvine \\
\\
Andrew Paverd & Gene Tsudik                      & Ai Enkoji\footnotemark{} \\
Microsoft     & UC Irvine &  LLNL  \\
\end{tabularx}
}        

        \newcommand{\shortauthors}{Searles et al.}

  	\maketitle
	
	\begin{abstract}
	{\input{content/00-abstract.tex}}
	\end{abstract}
	
	{\let\thefootnote\relax\footnote{{$^{*,\dagger}$~Work done while at UC Irvine.}}}

	\input{content/01-introduction.tex}

	\input{content/02-research_findings.tex}

	\input{content/03-internet_survey.tex}
	
	\input{content/04-user_study.tex}

	\input{content/05-evaluation.tex}

	\input{content/06-related_work.tex}

	\input{content/07-conclusion.tex}

	\input{content/09-acknowledgements.tex}

	\footnotesize
	\raggedbottom
	\bibliographystyle{abbrv}
	\bibliography{references.bib}

	\pagebreak
	\normalsize
	\appendix
	\pagebreak
	\input{content/10-appendix.tex}

\end{document}

%% file: content/00-abstract.tex
For nearly two decades, \captchas have been widely used as a means of protection against bots. 
Throughout the years, as their use grew, techniques to defeat or bypass \captchas have continued to improve.
Meanwhile, \captchas have also evolved in terms of sophistication and diversity, becoming increasingly difficult to solve for both bots (machines) and humans. 
Given this long-standing and still-ongoing arms race, it is critical to investigate how long it takes legitimate users to solve modern \captchas, and how they are perceived by those users.

In this work, we explore \captchas \emph{in the wild} by evaluating users' solving performance and perceptions of \emph{unmodified currently-deployed} \captchas.
We obtain this data through manual inspection of popular websites and user studies in which $1,400$ participants collectively solved $14,000$ \captchas.
Results show significant differences between the most popular types of \captchas: surprisingly, solving time and user perception are not always correlated.
We performed a comparative study to investigate the effect of \emph{experimental context} -- specifically the difference between solving \captchas directly versus solving them as part of a more natural task, such as account creation.
Whilst there were several potential confounding factors, our results show that experimental context could have an impact on this task, and must be taken into account in future \captcha studies.
Finally, we investigate \captcha-induced user task \emph{abandonment} by analyzing participants who start and do not complete the task.

%% file: content/01-introduction.tex
\section{Introduction}
\label{sec:intro}

Automated bots pose a significant challenge for, and danger to, many website operators and providers.
Masquerading as legitimate human users, these bots are often programmed to scrape content, create accounts, post fake comments or reviews, consume scarce resources, or generally (ab)use other website functionality intended for human use~\cite{imperva2022, forrester2021}.
If left unchecked, bots can perform these nefarious actions at scale.
\Captchas are a widely-deployed defense mechanism that aims to prevent bots from interacting with websites by forcing each user to perform a task, such as solving a challenge~\cite{cap_usage}.
Ideally, the task should be straightforward for humans, yet difficult for machines~\cite{vonAhn}.

The earliest \captchas asked users to transcribe random distorted text from an image.
However, advances in computer vision and machine learning have dramatically increased the ability of bots to recognize distorted text~\cite{yan2008low,gao2012divide,hernandez2010pitfalls}, and by 2014, automated tools achieved over 99\% accuracy~\cite{Goodfellow,Shet}.
Alternatively, bots often outsource solving to \emph{\captcha farms} -- sweatshop-like operations where humans are paid to solve \captchas~\cite{Motoyama}.
In light of this, \captchas have changed and evolved significantly over the years. 
Popular \captcha tasks currently include object recognition (e.g., ``select squares with...''), parsing distorted text, puzzle solving (e.g., ``slide the block...''), and user behavior analysis~\cite{Goodfellow,Shet}. 
It is therefore critical to understand and quantify how long it takes legitimate users to solve current \captchas, and how these \captchas are perceived by users.

Several prior research efforts have explored \captcha solving times, e.g.,~\cite{Bursztein, Bigham, Gao, Ross, Uzun, senCAP}.
For example, over a decade ago, Bursztein et al.~\cite{Bursztein} performed a large-scale user study, using over $1,100$ unique participants from Amazon Mechanical Turk (MTurk)~\cite{mturk} as well as \captcha farms.
Their results showed that \captchas were often more difficult or took longer to solve than was expected.
There was a loose correlation between time-to-annoyance and abandonment, with higher abandonment rates observed for \captchas that took longer to solve.
The same study also showed several demographic trends, e.g., users outside the US typically took longer to solve English-language \captcha schemes.
However, since this study, the \captcha ecosystem has changed substantially: new \captcha types emerged, input methods evolved, and Web use boomed.

More recently, Feng et al.~\cite{senCAP} used a similar methodology, with $202$ participants, to study the usability of their newly-proposed sen\captcha in comparison to text, audio, image, and video-based \captchas.
They found that sen\captcha outperformed the alternatives, both in terms of solving time and user preference.
They used Securimage~\cite{securimage}, a free open-source PHP script, to generate text and audio \captchas, and they implemented their own image and video \captchas.

Building upon and complementing prior work, this paper evaluates \captchas \emph{in the wild} -- specifically, the solving times and user perceptions of \emph{unmodified} (i.e., not re-implemented) \emph{currently-deployed} \captcha types.
We first performed a manual inspection of 200 popular websites, based on the Alexa Top websites list~\cite{alexa}, to ascertain: (1) \emph{how many} websites use \captchas, and (2) \emph{what types} of \captchas they use. 
Next, we conducted a $1,000$-participant user study using Amazon MTurk, wherein each participant was required to solve $10$ different types of \captchas.
We collected information about participants' \captcha solving times, relative preferences for \captcha types, types of devices used, and various demographic information.

One notable aspect of our user study is that we attempted to measure the impact of experimental context on participants' \captcha solving times.
Half of the participants were directly asked to solve \captchas, whilst the other half were asked to create accounts, which involved solving \captchas as part of the task.
The latter setting was designed to measure \captcha solving times \emph{in the context} of a typical web activity.  

One inherent limitation of any user study, especially when using MTurk, is that we cannot ensure that all participants who begin the study will complete it.
All of our results should therefore be interpreted as referring to \emph{users who are willing to solve \captchas}, rather than users in general.

Indeed, having noted that some participants began but did not complete our main study, we conducted a secondary MTurk study specifically designed to quantify how many users abandon their intended web activity when confronted with different types of \captchas. 
We believe that \captcha-induced \emph{user abandonment} is an important -- yet understudied -- consideration, since every abandoned task (e.g., purchase, account creation) represents a potential loss for the website.

To facilitate reproducibility and enable further analysis, we provide the entire anonymized data-set collected during our user studies, along with our analysis code.\footnote{\url{https://github.com/sprout-uci/captcha-study}}

%% file: content/02-research_findings.tex
\section{Research Questions \& Main Findings}
\label{sec:findings}

We now present our research questions and summarize our main findings.
Table~\ref{tab:findings} shows how our findings relate to prior work at a high level, with detailed comparisons in Section~\ref{sec:related}.

\textbf{RQ1: How long do human users take to solve different types of \captchas?} 
Specifically, we aimed to measure solving times for \captchas that users are likely to encounter (e.g., those used on popular websites).
Our results align with previous findings~\cite{Bursztein, senCAP, Bigham} in showing that there are significant differences in mean solving times between \captcha types.
For comparison, we also identified the current fastest attacks on each type of \captcha (Table~\ref{tab:bots_vs_humans}).

\textbf{RQ2: What \Captcha types do users prefer?}
In order to understand users' relative preference for various types of \captchas, we asked participants to rate all \captcha types on a Likert scale of $1-5$, from least to most enjoyable. 
Our results show that there are marked differences in participants' preferences, with average preference scores ranging from $2.76$ to $3.94$.
Our results also show that average solving time is \emph{not fully correlated} with participants' preferences, which means that other factors, beyond the amount of time required to solve a \captcha, influence participants' preferences.
Our analysis of data from prior studies~\cite{Krol, senCAP, tanth19} shows that their data supports this finding (even if they do not discuss it explicitly).

\textbf{RQ3: Does experimental context affect solving time?}
Specifically, we aimed to quantify the difference in solving times between the setting where participants are directly tasked with solving \captchas versus the setting in which participants solve \captchas as part of a typical web activity, such as user account creation.
We therefore ran two separate versions of our main user study: \emph{direct} and \emph{contextualized}, which we describe in detail in Section~\ref{sec:B_vs_UB_SD}.
Whilst there were several potential confounding factors in our study, our results show that experimental context could have an impact on \captcha user studies, with the difference in mean solving times as high as 57.5\% in our study.

\textbf{RQ4: Do demographics affect solving time?}
We analyzed different self-reported metrics including age, gender, country of residence, education, Internet usage, device type and input method.
In line with prior results~\cite{Bursztein}, we found that all types of \captchas take longer for older participants.
Specifically, \cite{Bursztein} reported an increase in solving time for text-based \captchas of $0.03$ seconds per year of participant age.
Our results show an even stronger dependence with an average increase across all \captcha types of $0.09$ seconds per year.
Additionally, \cite{Bursztein} showed that participants with a PhD solved \captchas faster than all other educational groups.
In contrast, our results show that our participants' self-reported level of education does not correlate with their solving times.

\textbf{RQ5: Does experimental context influence abandonment?}
Specifically, we aimed to quantify the extent to which abandonment within a \captcha user study is influenced by i) experimental context, and ii) the amount of compensation offered.
For different combinations of the above variables, we found that between 18\% and 45\% of participants abandoned the study after the presentation of the first \captcha.
Only one prior \captcha user study~\cite{Bursztein} disclosed their observed rate of abandonment, which is similar to that observed in our study.
Overall, participants in the contextualized setting were 120\% more likely to abandon than their peers in the direct setting.
This connection between experimental context and user abandonment is a new finding.

\begin{table*}[ht]
\footnotesize
\caption{Summary of research questions and main findings.}
\label{tab:findings}
\begin{tabularx}{\textwidth}{X X X X}
\toprule
& \textbf{Findings supporting prior work} 
& \textbf{Findings contradicting prior work} 
& \textbf{New findings on \Captchas} 
\\\toprule
\textbf{RQ1: How long does it take humans to solve different types of \captchas?} 
&  Solving time across \captcha types has a large degree of variance. \cite{Bursztein, senCAP, Bigham}
&                                              
&  
\\ \midrule
\textbf{RQ2: What \Captcha types do users prefer?}                         
&   Solving time is not correlated with user preference. \cite{Krol, senCAP, tanth19}                
&                                  
&  
\\ \midrule
\textbf{RQ3: Does experimental context affect solving time?}                         
&                                            
&                                               
&  Solving time is heavily influenced by experimental context, with differences in means up to 57.5\%.
\\ \midrule
\textbf{RQ4: Do demographics affect solving time?}                                      
&  Age has an effect on solving time. \cite{Bursztein}
&  Self-reported education does not correlate with solving time. \cite{Bursztein}
&  %
\\ \midrule
\textbf{RQ5: Does experimental context influence abandonment?}                        
&   High abandonment rates observed in \captcha user studies. \cite{Bursztein}
&                                               
&   Experimental context directly affects the rate of abandonment. 
\\ \bottomrule                             
\end{tabularx}
\end{table*}

%% file: content/03-internet_survey.tex
\section{Website Inspection} 
\label{sec:inspection}
To understand the landscape of modern \captchas and guide the design of the subsequent user study, we manually inspected the 200 most popular websites from the Alexa Top Website list~\cite{alexa}.
Where applicable, we use the terminology from the taxonomy proposed by Guerar et al.~\cite{Guerar}.

Our goal was to imitate a normal user's web experience and trigger \captchas in a natural setting.
Although \captchas can be used to protect any section or action on a website, they are often encountered during user account creation to prevent bots creating accounts.
Thus, for each website, we investigated the process of creating an account (wherever available).
Of the inspected websites, $185$ had some type of account creation process, and we could successfully create accounts on $142$ websites.
Distinct domains operated by the same organization (e.g., \url{amazon.com} and \url{amazon.co.jp}) were counted separately.
We visited each website twice: once with Google Chrome in incognito mode, and once with the Tor browser over the Tor network~\cite{tor}.
We used incognito mode to avoid websites changing their behavior based on cookies presented by our browser.
We used Tor since anecdotal evidence suggests Tor users are asked to solve \captchas more frequently and with greater difficulty than non-Tor users.
If no \captchas were displayed, we searched the page source for the string ``\captcha'' (case insensitive).

\taggedpara{Ethical considerations:} Based on the Guidelines for Internet Measurement Activities~\cite{rfc1262}, we did not engage in malicious behavior which may trigger additional \captchas.
We used only manual analysis to avoid various challenges that arise from automated website crawling.

\subsection{Results and analysis}
\label{subsec:inspection_results}
\begin{figure}[t!]
	\centering
	\includegraphics[width=\linewidth]{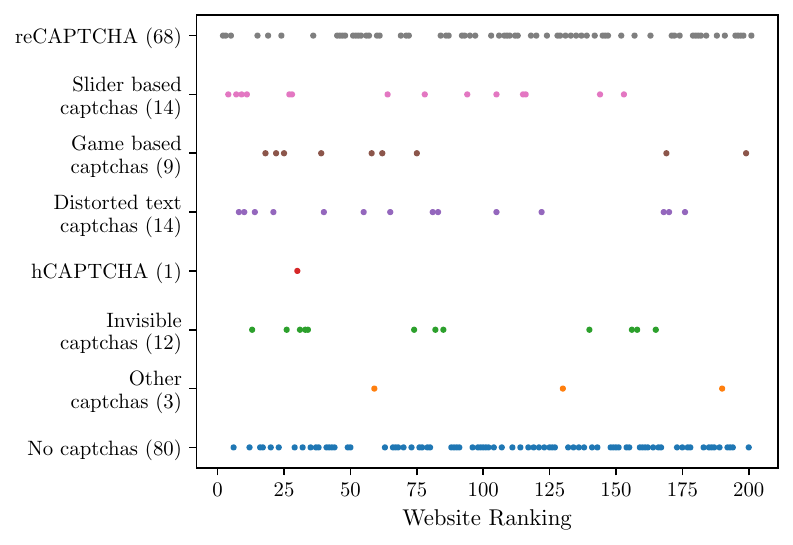}
	\caption{Discrete distribution of discovered \captchas (full data available in the accompanying dataset).}
	\label{fig:disc_cap}
\end{figure}

Figure~\ref{fig:disc_cap} shows the distribution of \captcha types we observed during our inspection.
The most prevalent types were:

\taggedpara{reCAPTCHA}~\cite{reCAPTCHAv2,reCAPTCHA,reCAPTCHAv3} was the most prevalent, appearing on 68 websites (34\% of the inspected websites).
This is a Google-owned and operated service that presents users with ``click'' tasks, which include behavioral analytics and may potentially result in an image challenge.
reCAPTCHA allows website operators to select a difficulty level, ranging from ``easiest for users'' to ``most secure''.

\taggedpara{Slider-based} \captchas appeared on 14 websites (7\%).
These typically ask users to slide a puzzle piece into a corresponding empty spot using a drag interaction.
The timing and accuracy is checked for bot-like behavior.

\taggedpara{Distorted Text} \captchas appeared on 14 websites (7\%).
We observed differences in terms of text type, color, length, masking, spacing, movement, and background.
Text type varied in several ways: 2D or 3D, solid or hollow, font, and level of distortion.
Certain \captchas used masking, i.e., lines or shapes obscured parts of the letters.

\taggedpara{Game-based} \captchas appeared on 9 websites (4.5\%).
These present users with dynamic games and compute a risk profile from the results.
For example, users are asked to rotate an image or select the correctly oriented image.

\taggedpara{hCAPTCHA}~\cite{hCaptcha} appeared on 1 website.
This is a service provided by Intuition Machines, Inc.\ that was recently adopted by Cloudflare~\cite{cloudflare} and is gaining popularity.

\taggedpara{Invisible \captchas} were found on 12 websites (6\%).
These websites did not display any visible \captchas, but contained the string ``\captcha'' in the page source.

\taggedpara{Other \Captcha{s}} found during our inspection included: a \captcha resembling a scratch-off lottery ticket; a \captcha asking 
users to locate Chinese characters within an image; and a proprietary \captcha service called ``NuCaptcha''~\cite{nucaptcha}.

\subsection{Potential limitations}
\label{sec:inspection_limitations}

\taggedpara{Choice of website list:}
There are several lists of \emph{``popular''} websites that could be used for this type of study, including the Alexa Top Website list~\cite{alexa}, Cisco Umbrella~\cite{ciscoUmbrella}, Majestic~\cite{majestic}, TRANCO~\cite{pochat2019tranco}, Cloudflare Radar~\cite{cloudflare_radar}, and SecRank TopDomain~\cite{xie2022secrank}.
These lists vary because of the differences in the methodology used to identify and rank websites.
Following the work of Bursztein et al.~\cite{Bursztein} and the recommendation of Scheitle et al.~\cite{Scheitle2018}, we used the Alexa list.

\taggedpara{Number of inspected websites:}
Since our website inspection was a manual process, we could only inspect the top 200 websites.
This may also introduce a degree of systemic bias towards the types of \captchas used on the most popular websites.
However, we specifically chose these websites because they are visited by large numbers of users.

\taggedpara{Lower bound:}
Since we did not exercise all possible functionality of every website, it is possible that we might not have encountered all \captchas. 
Therefore, our results represent a lower bound, while the actual number of deployed \captchas may be higher.
Nevertheless, we believe that we identified the most prevalent \captcha types across all inspected websites.

\taggedpara{Timing:}
Web page rankings change on the daily basis and \captcha{s} shown by the same service may change.
Given that our inspection was performed at a particular point in time, the precise results will likely change if the inspection were repeated at a different point in time.
However, as explained above, we believe that the identified set of \captcha types is representative of currently-deployed \captchas.

\taggedpara{Other types of \captchas:}
We only inspected mainstream websites (i.e., those that would appear on a top websites list).
This means that there could be \captchas that are prevalent on other types of websites (e.g., on the dark web) but are not included in our study.
However, studying these \emph{special-purpose} \captchas might require recruiting participants who have prior experience solving them, which was beyond the scope of our study.

\taggedpara{Impact of limitations:}
The above limitations could have had an impact on the set of \captcha types we identified and subsequently used in our user study.
However, we have high confidence that the \captcha types we identified are a realistic sample of those a real user would encounter during typical web browsing.
For instance, BuiltWith~\cite{cap_usage} has analyzed a dataset of 673 million websites and identified 15.2 million websites that use \captchas.
reCAPTCHA accounts for 97.3\% and hCAPTCHA for a further 1.4\%.
The \captcha types used in our study therefore account for over 98\% of \captchas in this large-scale dataset.

%% file: content/04-user_study.tex
\section{User Study} \label{sec:user_study}

Having identified the relevant \captcha types, we conducted a $1,000$ participant online user study to evaluate real users' solving times and preferences for these types of \captchas.
Our study was run using using Amazon MTurk and can be summarized into the following four phases:

\taggedpara{1. Introduction:} Participants were first given an overview of the study and details of the tasks to complete.

\taggedpara{2. Pre-study questions:} All participants were then asked to provide demographic information by answering the pre-study questions shown in Table~\ref{tab:questions} in Appendix~\ref{sec:appendix-questions}.

\taggedpara{3. Tasks:} Participants were asked to complete tasks, which included solving exactly ten \captchas, presented in random order.
Unless otherwise stated, each \captcha was \emph{unique} (i.e., freshly generated per participant).
Participants had to solve each \captcha in order to progress to the next step, thus preventing them from speeding through the study.

\taggedpara{4. Post-study question} Finally, participants were asked questions about the \captchas they had just solved.
The exact questions and possible answers are shown in Table~\ref{tab:questions} in Appendix~\ref{sec:appendix-questions}.

\begin{figure*}[!t]
    \centering
    \begin{minipage}{.33\textwidth}
        \centering
        \begin{subfigure}[t]{\linewidth}
			\centering
			\includegraphics[width=0.7\linewidth]{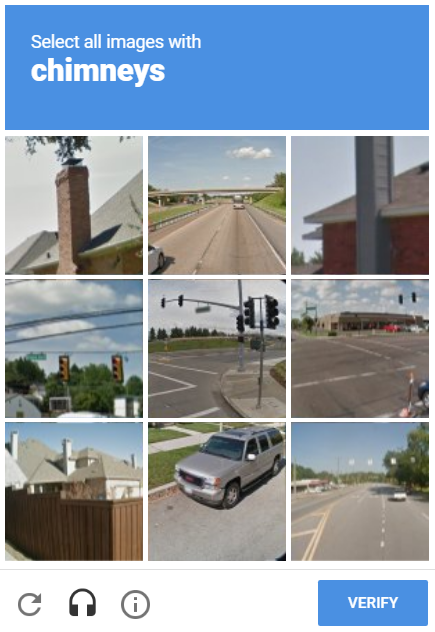}
			\caption{Image Task \captcha~\cite{reCAPTCHA}}
			\label{fig:recaptcha_image}
		\end{subfigure}\vspace{15mm}
		\begin{subfigure}[b]{\linewidth}
			\centering
			\includegraphics[width=0.9\linewidth]{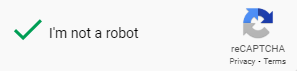}
			\caption{v2 checkbox \captcha~\cite{reCAPTCHAv2}}
			\label{fig:recaptcha_check}
		\end{subfigure}
		\label{fig:recaptcha}
		\caption{reCAPTCHA~\cite{reCAPTCHAv2,reCAPTCHA,reCAPTCHAv3}}
    \end{minipage}%
    \begin{minipage}{0.33\textwidth}
        \centering
        \begin{subfigure}[t]{\linewidth}
			\centering
			\includegraphics[width=0.7\linewidth]{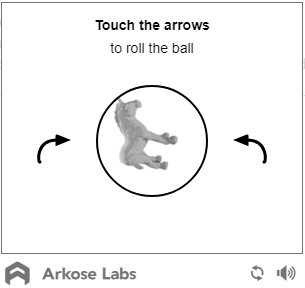}
			\caption{Rotation \captcha}
			\label{fig:arkose_rotate}
		\end{subfigure}
		\begin{subfigure}[b]{0.7\linewidth}
			\centering
			\includegraphics[trim=0 0 0 0, clip, width=\linewidth]{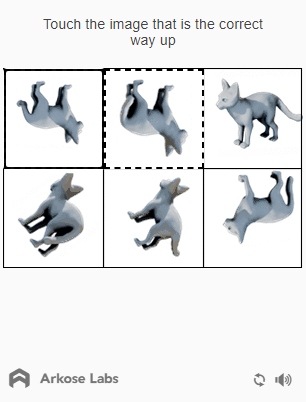}
			\caption{Orientation selection}
			\label{fig:arkose_orientation}
		\end{subfigure}
		\label{fig:arkose}
		\caption{Arkose Labs~\cite{arkose}}
    \end{minipage}%
    \begin{minipage}{0.33\textwidth}
        \centering
		\includegraphics[trim=0 10mm 0 0, clip, width=0.7\linewidth]{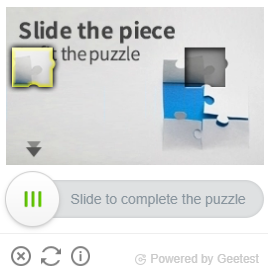}
		\caption{Geetest~\cite{geetest}}
		\label{fig:geetest}
		\includegraphics[trim=0 0 0 0, clip, width=0.7\linewidth]{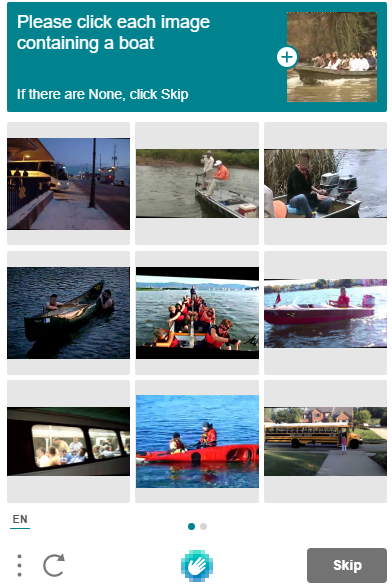}
		\caption{hCAPTCHA~\cite{hCaptcha}}
		\label{fig:hcaptcha}
    \end{minipage}
\end{figure*}

\begin{figure}[t]
	\centering
	\begin{minipage}[b]{0.45\linewidth}
		\begin{subfigure}[b]{\linewidth}
			\centering
			\includegraphics[width=0.8\linewidth]{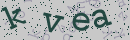}
			\caption{Xinhuanet \captcha~\cite{Xinhuanet}}
			\label{fig:Xinhuanet}
		\end{subfigure}
		\begin{subfigure}[b]{\linewidth}
			\centering
			\includegraphics[width=0.8\linewidth]{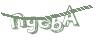}
			\caption{360.cn \captcha~\cite{360}}
			\label{fig:360}
		\end{subfigure}
	\end{minipage}
		\begin{subfigure}[b]{0.45\linewidth}
			\centering
			\includegraphics[width=0.45\linewidth]{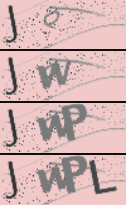}
			\caption{jrj.com \captcha~\cite{jrj}}
			\label{fig:jrj}
		\end{subfigure}
	\caption{Distorted text \captchas}
	\label{fig:distorted_text_captcha}
\end{figure}

\subsection{Choice of \Captchas}
\label{sec:choice_of_captchas}
Based on our website inspection (Section~\ref{sec:inspection}), we selected the following ten types of \captchas:
\begin{itemize}[itemsep=.2mm]	
	\item Two reCAPTCHA v2 \captchas: one with the setting \emph{easiest for users} and the other with \emph{most secure}. Note that we do not have control over whether the user is shown an image-based (Figure~\ref{fig:recaptcha_image}) challenge in addition to the click-based (Figure~\ref{fig:recaptcha_check}) task.
	\item Two game-based \captchas from Arkose Labs~\cite{arkose}: one required using arrows to rotate an object (Figure~\ref{fig:arkose_rotate}) and the other required selecting the upright object (Figure~\ref{fig:arkose_orientation}).
	\item Two hCAPTCHAs~\cite{hCaptcha}: one with easy and one with difficult settings (Figure~\ref{fig:hcaptcha}). 
	\item One slider-based \captcha from Geetest~\cite{geetest}: we selected Geetest because it was used on several of the inspected websites and offers a convenient API (Figure~\ref{fig:geetest}).
	\item Three types of distorted text \captchas (Figure~\ref{fig:distorted_text_captcha}): (a) the \emph{simple} version had four unobscured characters, (b) the \emph{masked} version had five characters and included some masking effects, and (c) the \emph{moving} version contained moving characters.
\end{itemize}
These form a representative sample of \captchas we encountered in our website inspection.
Although hCAPTCHA only appeared once, we included it since it is an emerging image-based approach, which claims to be the largest independent \captcha service~\cite{hCaptcha_largest}.

\subsection{Direct vs. contextualized settings}
\label{sec:B_vs_UB_SD}
We initially hypothesized that we would observe a difference in behavior depending on experimental context.
In order to evaluate this, we designed two settings of the study: 500 participants completed the \emph{direct setting}, whilst the other 
500 completed the \emph{contextualized setting}.
In both settings, each participant solved exactly ten \captchas in random order. %

\taggedpara{Direct setting:} This setting was designed to match previous \captcha user studies, in which participants are directly asked to solve \captchas.
The MTurk study title was ``CAPTCHA User Study'' and the instructions in the first phase informed users that their task was to solve \captchas.
In the second phase, in addition to the basic demographic information, participants were asked about their experience with and perception of \captchas; see Table~\ref{tab:questions} in Appendix~\ref{sec:appendix-questions}.
In the third phase, participants were shown ten \captchas in random order. %
The fourth phase was the same for both settings.

\taggedpara{Contextualized setting:} This setting was designed to measure \captcha solving behavior \emph{in the context} of a typical web activity.
We selected the task of user account creation, as this often includes solving a \captcha. 
The MTurk study title was ``Account Creation User Study'' and the first and second phases did not mention \captchas.
In the third phase, participants were asked to complete ten typical user account creation forms, each displaying a \captcha \emph{after} the participant clicked submit, as is often the case on real websites.
This sequencing allowed us to precisely measure the \captcha solving time in isolation from the rest of the account creation task. 
The account creation task was a basic web form asking for a randomized subset of: name, email address, phone number, password, and address.
To avoid collecting personally identifiable information, participants were provided with synthetic information at each step.
Each page also included a large banner clearly stating not to enter any personal information.
The fact that we were specifically measuring \captcha solving time was only revealed to participants after they completed the first three phases.

\begin{table*}[tb]
\footnotesize
\caption{Summary of demographic data for the $1,400$ participants of the main user study.}
\label{tab:demographics}
	\begin{tabularx}{\textwidth}[t]{X|X|X|X|X|X|X}
	\toprule
		\textbf{Age} & \textbf{Residence} & \textbf{Education} & \textbf{Gender} & \textbf{Device Type} & \textbf{Input Method} & \textbf{Internet Use} \\
		\cmidrule(r){1-1} \cmidrule(lr){2-2} \cmidrule(lr){3-3} \cmidrule(lr){4-4} \cmidrule(lr){5-5} \cmidrule(lr){6-6} \cmidrule(l){7-7}
		30 - 39 (531) & USA (985) & Bachelors (822) & Male (832) & Computer (1301) & Keyboard (1261) & Work (860)\\
		20 - 29 (403) & India (240) & Masters (243) & Female (557) & Phone (74) & Touch (125) & Web surf (397)\\
		40 - 49 (271) & Brazil (50) & High school (210) & Non-Binary (11) & Tablet (25) & Other (14) & Education (87)\\
		50 - 59 (106)  & Italy (27) & Associate (98) & \multirow{3}{*}{} & \multirow{3}{*}{} & \multirow{3}{*}{} & Gaming (30) \\
		\;\;\;$\geq 60$ (58)     & UK (24) & PhD (24) & & & & Other (26)\\
		18 - 19 (31)  & Other (74) & No degree (3) & & & & \\
	\bottomrule
	\end{tabularx}
\end{table*}

\subsection{Timeline and compensation}
\label{sec:time_and_comp}
The primary study ran for two months with a total of $1,000$ distinct participants.\footnote{To the best of our knowledge, all participants were distinct. We configured Amazon MTurk to only allow unique accounts to participate.}
Participants were initially paid \$0.30 for completing the direct version and \$0.75 for the contextualized version, as the latter involved a larger workload. 
After completing the study, we realized we may have unintentionally under-compensated participants,\footnote{In terms of US federal minimum wage.} since the median HIT completion time was $4.4$ and $11.5$ minutes for direct and contextualized versions.
We therefore retroactively doubled all participants' compensation to $\$0.60$ and $\$1.50$, which equates to approximately $\$7.80$ -- $\$8.20$ per hour.

\subsection{Ethical considerations}
\label{sec:ethical}
This user study was duly approved by the Institutional Review Board (IRB) of the primary authors' organization.
No sensitive or personally identifiable information was collected from participants. 
We used the pseudonymous MTurk worker IDs only to check that participants were unique.

Since the contextualized setting did not inform participants of the actual aim of the study beforehand, two additional documents were filed and approved by the IRB: 
(1) \emph{``Use of deception/incomplete disclosure''} and (2) \emph{``Waiver or Alteration of the Consent''}.
After each participant completed the contextualized setting, we disclosed the study's actual goal and asked whether they gave us permission to use their data.
No data were collected from participants who declined.

\subsection{User study implementation}
\label{sec:user_study_impl}
The realization of the user study included a front-end webpage and a back-end server.
The front-end was a single HTML page that implemented the four phases described above.
To prevent any inconsistencies, participants were prevented from going back to a previous phase or retrying a task once they had progressed.
Timing events were captured with millisecond precision using the native JavaScript \texttt{Date} library.
Timing events were recorded at several points for each \captcha: request, serve, load, display, submit, and server response.
We measured \emph{solving time} as the time between a \captcha being displayed and the participant submitting a solution, as is done in prior \captcha user studies~\cite{Bursztein,Bigham,fidas2011,yan2008,belk,Ho,Krol,Ross,Uzun,Gao,mohitAutomatic2019,manarDynamic2014,gaoEmerging2019}.
Depending on the type of \captcha, this might include multiple rounds or attempts.

We used Amazon MTurk to recruit participants, host the front-end, and collect data.
While most types of \captchas shown by the front-end were served from their respective providers, distorted text \captchas were not available from a third-party provider, as these are usually hosted by the websites themselves.
We therefore set up our own back-end server to serve distorted text \captchas. 
Specifically, we downloaded a total of 1,000 unique distorted text \captchas of three different types, and stored these in a local \texttt{MongoDB}~\cite{mongodb} database.
We used a \texttt{Node.js}~\cite{nodejs} server to retrieve and serve \captchas from the database.
Every participant was served one text \captcha of each type, and each unique text \captcha was served to three different participants.

Table~\ref{tab:demographics} shows the demographic information of the participants who completed the study.
The demographics of the two subgroups who completed direct and contextualized studies are very similar to each other.

\subsection{Potential limitations}
\label{sec:study_limitations}
\taggedpara{Use of MTurk:}
Webb et al.~\cite{webb2022too} reported several potential concerns regarding the quality of data collected from MTurk.
Of their six criteria, our study did not implement two: consent quiz (1) and examination of qualitative responses (2), which we acknowledge as a limitation.
The remaining four criteria can be either evaluated through collected data or are not an issue for our study.
Eligibility (3) and attention check (4) can be verified via the accuracy of text-based \captcha responses, which confirm that nearly all of our participants were focused and provided correct data.
Response time (5) was within our expected range.
Study completion (6) was not an issue, since each participant had to complete every \captcha to proceed.

\taggedpara{Bots and farms:}
Similarly, Chmielewski et al.~\cite{chmielewski2020mturk} reported a decrease in data quality, citing bot and farm activity.
However, Moss and Litman~\cite{moss_2020} subsequently used several bot-detection measures to evaluate whether bots could be contaminating MTurk data, and found no evidence of bot activity.
Every participant who completed our study solved ten modern \captchas, which although possible, would be more difficult for bots.
Since we configured MTurk to only allow one completion per MTurk account, farm activity was also limited.
Therefore, we are reasonably confident that our results are not influenced by bots or farms.

\taggedpara{Choice of \captchas:}
One consequence of using the \captcha types we identified in Section~\ref{sec:inspection} is that our user study results are not directly comparable with those from prior \captcha user studies.
In general, it is difficult to directly compare such studies, as even if the same \emph{types} of \captchas are studied, different implementations may be used e.g., reCAPTCHA and hCAPTCHA are both image-based \captchas, but could give different results.

\taggedpara{Unmodified \captchas:}
In order to maximize the level of realism in our study, we used existing unmodified \captchas. 
We therefore did not have fine-grained control over the precise behavior of these \captchas, nor the ability to obtain more fine-grained measurements of participants' accuracy or performance beyond overall solving time.
However, like previous studies, we consider overall solving time to be the most important measurable quantity.

\taggedpara{Invalid inputs:}
Unfortunately, the input field for the \captcha preference question in our post-study questionnaire was a free text field rather than a pull-down menu.
This allowed some participants to provide preference scores outside the requested 1-5 range.
We therefore excluded invalid preference scores from 163 participants.\footnote{However, we have high confidence that these participants did not provide incorrect or rushed responses during the rest of the study because their average accuracy in text-based \captchas was similar to the study-wide average.
We therefore retained their measurements in other sections.}

\taggedpara{Abandonment:}
Since we did not record how many participants began our main study, we cannot precisely quantify the rate of abandonment.
To investigate this further, we performed an additional abandonment-focused study (Section~\ref{sec:eval_abandonment}), where we observed a $30\%$ abandonment rate.
We can therefore assume a similar abandonment rate for our main study.
Whilst the impact of this level of abandonment is unclear, it could potentially affect the ecological validity of our results, as the participants who were willing to complete the study may not be representative of all users.

\taggedpara{Confounding factors:}
There were several differences between our direct and contextualized settings, some of which may be confounding factors when comparing these two groups.
For example, participants in the contextualized setting had to do more work, so their attention or focus might have been reduced during \captcha solving.
Differences in compensation or participants' perceived benefit of completing the task (i.e., creating an account vs.\ solving a \captcha) may have affected motivation or likeliness to abandon the task.

%% file: content/05-evaluation.tex
\section{Results \& Analysis} 
\label{sec:eval}
This section presents the user study results.
Unless otherwise indicated, results are based on the full set of participants.

\subsection{Solving times}
\label{sec:eval_solving_time}
\begin{figure}[t]
		\centering
		\includegraphics[width=\columnwidth,trim=2mm 2mm 2mm 2mm,clip]{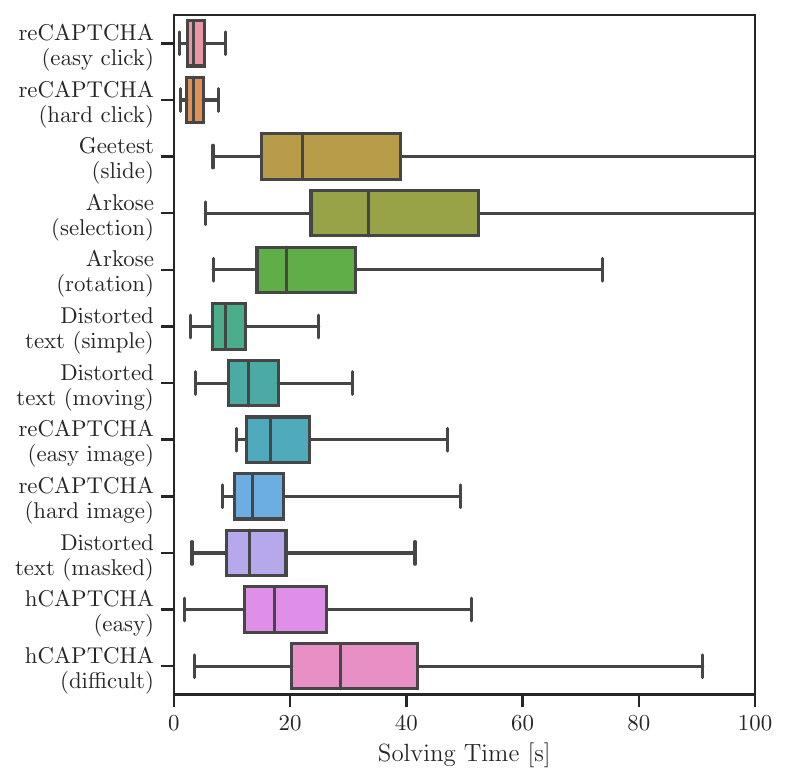}
		\caption{Solving times for various types of \captchas. Boxes show the middle 50\% of 
		participants, and whiskers show the filtered range. Black vertical lines show the median.}
		\label{fig:captcha_solving_time}
\end{figure}

This subsection addresses \textbf{RQ1:} \emph{How long do human users take to solve different types of \captchas?}
Figure~\ref{fig:captcha_solving_time} shows the the distribution of solving times for each \captcha type.
We observed a small number of extreme outliers where the participant likely switched to another task before returning to the study.
We therefore filtered out the highest $50$ solving times per \captcha type, out of $1,000$ total. %

For reCAPTCHA, the selection between image- or click-based tasks is made dynamically by Google.
Whilst we know that $85\%$ and $71\%$ of participants (easy and hard setting) were shown a click-based \captcha, the exact task-to-participant mapping is not revealed to website operators.
We therefore assume that the slowest solving times correspond to image-based tasks.
\todo{This seems to be entirely conjecture. Can you clarify this process. A comment on HotCRP is fine.}
After disambiguation, click-based reCAPTCHA had the lowest median solving time at $3.7$ seconds.
Curiously, there was little difference between easy and difficult settings.

The next lowest median solving times were for distorted text \captchas.
As expected, simple distorted text \captchas were solved the fastest.
Masked and moving versions had very similar solving times.
For hCAPTCHA, there is a clear distinction between easy and difficult settings.
The latter consistently served either a harder image-based task or increased the number of rounds.
However, for both hCAPTCHA settings, the fastest solving times are similar to those of reCAPTCHA and distorted text.
Finally, the game-based and slider-based \captchas generally yielded higher median solving times, though some participants still solved these relatively quickly (e.g., $< 10$~seconds).

With the exception of reCAPTCHA (click) and distorted text, we observed that solving times for other types have a relatively high variance. 
Some variance is expected, especially since these results encompass all input modalities across both direct and contextualized settings. 
However, \emph{relative differences in variances} indicate that, while some types of \captchas are consistently solved quickly, most have a range of solving times across the user population.
The full statistical analysis of our solving time results is presented in Appendix~\ref{sec:eval_statistical}.

\todo{I think a violin plot would be preferable to the box plots.}

\begin{figure}[t]
		\centering
		\includegraphics[width=\columnwidth,trim=2mm 2mm 2mm 2mm,clip]{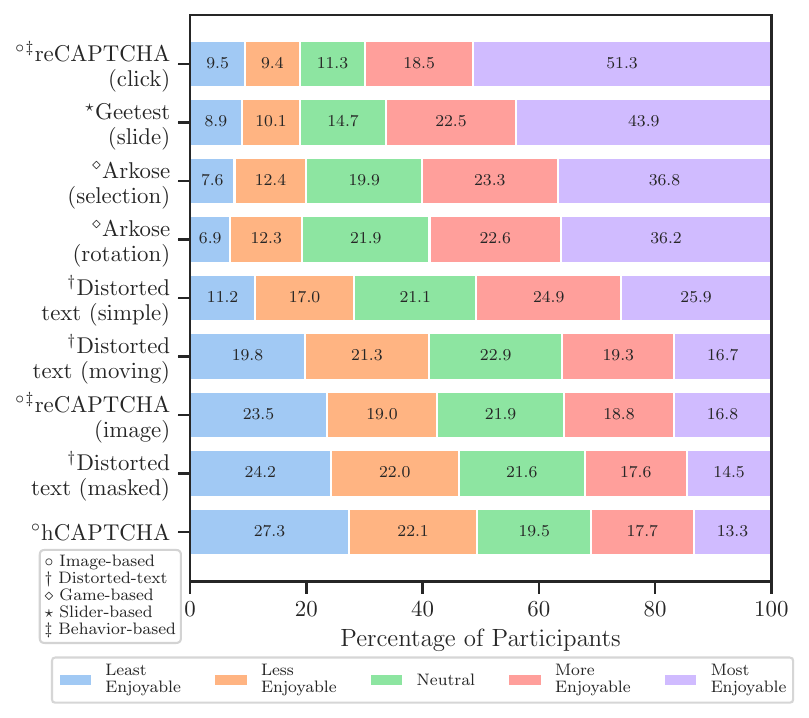}
		\caption{Participant-reported preference scores for different types of \captchas, 
		sorted from highest to lowest.}
		\label{fig:pref_res}
\end{figure}

\subsection{Preferences analysis}
\label{sec:eval_preferences}
This subsection addresses \textbf{RQ2:} \emph{What \Captcha types do users prefer?}
Figure~\ref{fig:pref_res} shows participants' \captcha preference responses after completing the solving tasks.
The \captcha types are sorted from most to least preferred by overall preference score, which is calculated by summing the numeric scores.
Since easy and difficult settings of hCAPTCHA are visually indistinguishable, we could only ask participants for one preference.

As expected, participants tend to prefer \captchas with lower solving times.
For example, reCAPTCHA (click) has the lowest median solving time and the highest user preference.
However, surprisingly, this trend does not seem to hold for game-based and slider-based \captchas, since these received some of the highest preference scores, even though they typically took longer than other types.
\todo{If there is room, it would be interesting to see a linear regression of these items per captcha.}
This suggests that factors beyond solving time could be contributing to participants' preference scores.
Notably, no single \captcha type is either universally liked or disliked.
For example, even the top-rated click-based reCAPTCHA, was rated 1 or 2 by $18.9\%$ of participants.
Similarly, over $31.0\%$ rated hCAPTCHA 4 or 5, although it had the lowest overall preference score.

\begin{figure*}[t!]
	\centering
	\includegraphics[,width=\textwidth]{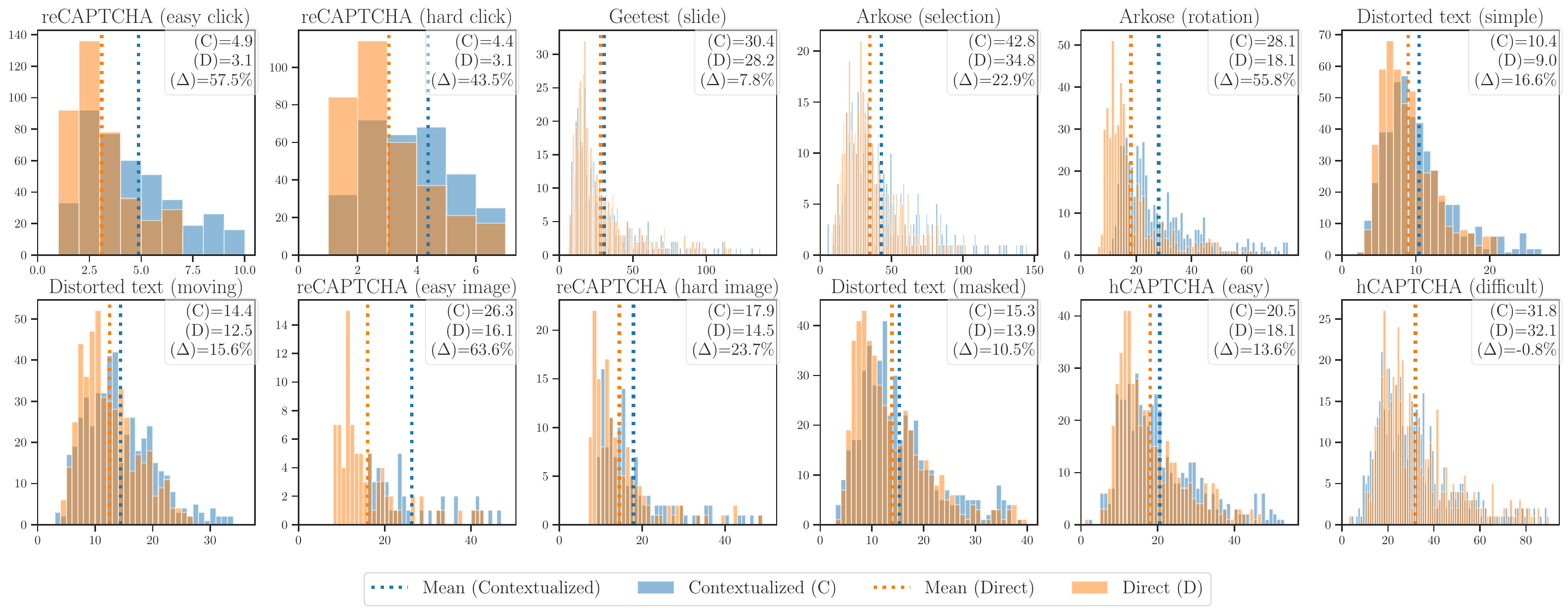}
	\caption{{\captcha solving times for direct (D) vs.\ contextualized (C) user study settings. 
	The horizontal axis shows solving time in seconds, quantized into one-second buckets, and the vertical axis shows number of participants.}}
	\label{fig:b_v_ub}
	\vspace{\floatsep}
	\includegraphics[width=\textwidth]{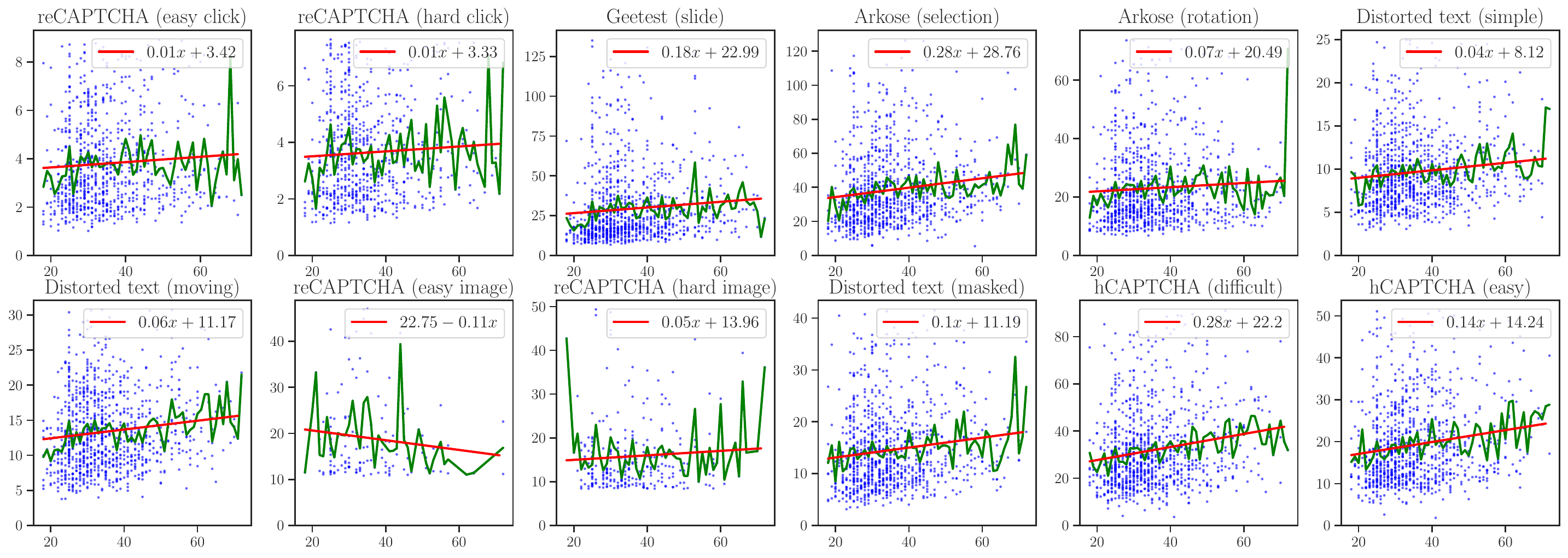}
    \caption{Effects of age in \captcha solving time. The horizontal axis shows the age and the vertical axis shows the solving time. The red line shows the linear fit of the data points and the green line shows the average solving time per age.}
    \label{fig:age_vs_time}
\end{figure*}

\subsection{Direct vs.\ contextualized setting}
\label{sec:eval_context}
This subsection addresses \textbf{RQ3:} \emph{Does experimental context affect solving time?}
Figure~\ref{fig:b_v_ub} shows histograms of \captcha solving times for participants in the direct vs.\ contextualized settings.
In every case except one, the mean solving time is lower in the direct setting.
In most cases, the distribution from the contextualized setting has more participants with longer solving times, i.e., a longer tail.

The largest statistically significant difference is in reCAPTCHA (easy click), where the mean solving time grows by $1.8$ seconds ($57.5\%$).
Second is Arkose (rotation), where it grows by $10$ seconds ($56.1\%$).
Across all \captcha types, the average increase from direct to contextualized is $26.7\%$.
Similarly, the mean solving time for reCAPTCHA (easy image) increased by $63.6\%$ in the contextualized setting showing the largest increase. 
However this was not statistically significant.
This is likely due to the skew of participants in direct and contextualized versions receiving image-challenges, which is controlled by Google.
Easy images were shown to 8.9\% of contextualized and to 17.2\% of direct setting participants, while hard images were shown to $25.5\%$ and $30\%$ respectively, resulting in different population sizes.

On the other hand, hCAPTCHA (difficult), which has the highest median solving time overall, showed no significant difference in mean solving time between direct and contextualized settings.
This may be attributable to the difficulty of solving this type of \captcha, regardless of the setting.

Results of Kruskal-Wallis tests confirm that there are statistically significant differences in mean solving times for all \captcha types ($p < 0.001$) except Geetest, reCAPTCHA (image) and hCAPTCHA (difficult).
While there were several potential confounding factors in our study, these results suggest that experimental context can have a significant impact on participants' \captcha solving times, and must therefore be taken into account in the design of future user studies.

\subsection{Effects of demographics}
\label{sec:eval_dem}
This subsection addresses \textbf{RQ4:} \emph{Do demographics affect solving time?}
We analyzed how demographic characteristics in our study correlate with \captcha solving times.
For some characteristics, such as education and gender, we did not observe large differences in \captcha solving times (see Figures~\ref{fig:gender_vs_time} and~\ref{fig:education_vs_time} in Appendix~\ref{sec:dem_appendix}).

\subsubsection{Effects of age}
\label{sec:eval_age}

Figure~\ref{fig:age_vs_time} shows the effect of participants' age on solving time.
The green line is the average solving time for each age, and the red line is a linear fit minimizing mean square error.
For all types, except reCAPTCHA (easy image), there is a trend of younger participants having lower average solving times.
This agrees with prior results~\cite{Bursztein} and is especially noticeable in hCAPTCHA, Arkose (selection), and Geetest.

\subsubsection{Effects of device type}
\label{sec:eval_device_type}
\begin{figure}[ht]
	\centering
	\includegraphics[width=\columnwidth,trim=0mm 2mm 0 1mm,clip]{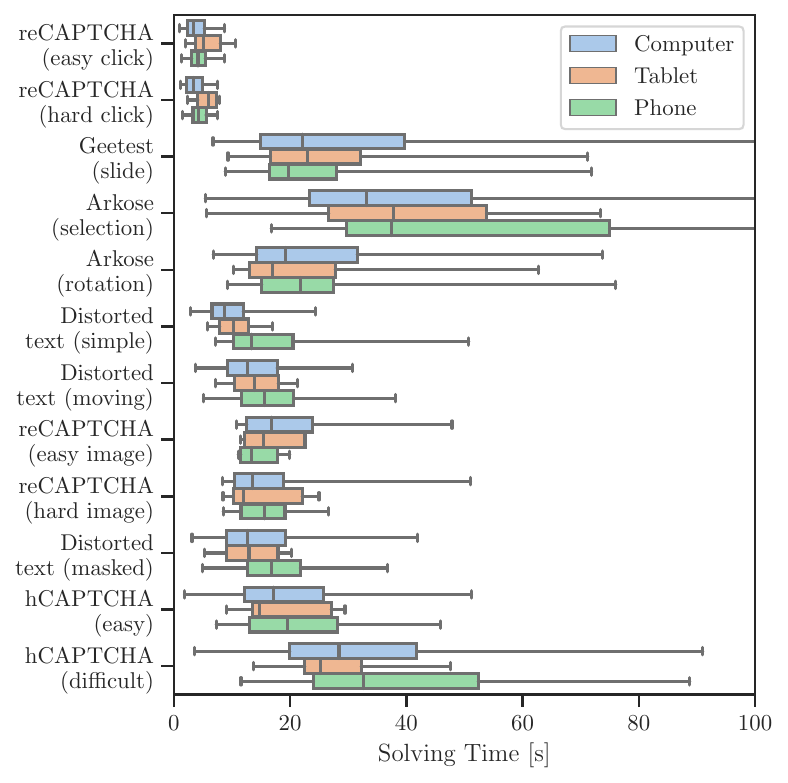}
	\caption{Effects of device type.}
	\label{fig:device_type_vs_time}
\end{figure}

Figure~\ref{fig:device_type_vs_time} shows the effect of device type.
Although there are some differences in median between device types for certain \captcha types, the Kruskal-Wallis test shows that the differences in means are mostly not statistically significant.
The only statistically significant differences are in distorted text \captchas ($p < 0.02$) and reCAPTCHA (hard click) ($p < 0.01$), where participants who used computers had a lower mean solving time compared to those using phones.
Interestingly, we found a statistically significant difference between participants who used physical keyboards and those who used touch input for the simple and masked distorted text \captchas ($p < 0.02$), as well as reCAPTCHA (hard click) ($p < .001$), reCAPTCHA (easy click) ($p < .05$), and Arkose (selection) ($p < .003$).
We found no statistically significant difference in mean solving times for moving distorted text. %

\subsubsection{Effects of typical Internet use}
\begin{figure}[ht]
	\includegraphics[width=\columnwidth,trim=0mm 2mm 0 1mm,clip]{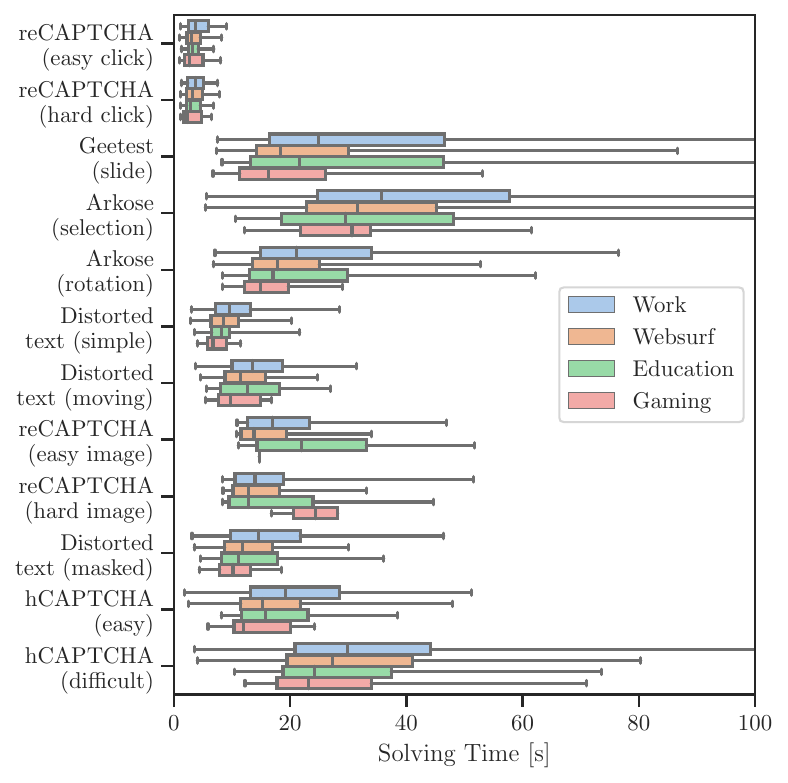}
	\caption{Effects of typical Internet use.}
	\label{fig:internet_use_vs_time}
\end{figure}

Figure~\ref{fig:internet_use_vs_time} shows the relationship between participants' self-reported dominant Internet usage patterns and their \captcha solving times. 
The Kruskal-Wallis test shows some initial evidence for statistically significant differences between participants who use the Internet primarily for work and those who use it for other purposes ($p < 0.05$).
The former were typically slower than the latter in 8 out of 12 \captchas. %
However, some categories do not have a sufficient number of participants, thus further investigation is recommended.

\subsection{Accuracy of \Captchas}
\label{sec:eval_accuracy}
Table~\ref{tab:bots_vs_humans} contrasts our measured human solving times and accuracy against those of automated bots reported in the literature.
Interestingly, these results suggest that bots \emph{can} outperform humans, both in terms of solving time and accuracy, across all these \captcha types.
As mentioned in Section~\ref{sec:study_limitations}, our decision to use unmodified real-world \captchas means we only have accuracy results for a subset of \captcha types (e.g., neither Geetest nor Arkose provide accuracy information).
For the same reason, our accuracy results also include participants who only partially completed the study.

\taggedpara{reCAPTCHA:} The accuracy of image classification was 81\% and 81.7\% on the easy and hard settings respectively.
Surprisingly, the difficulty appeared not to impact accuracy.

\taggedpara{hCAPTCHA:} The accuracy was 81.4\% and 70.6\% on the easy and hard settings respectively.
This shows that, unlike reCAPTCHA, the difficulty has a direct impact on accuracy.

\taggedpara{Distorted Text:} We evaluated \emph{agreement} among participants as a proxy for accuracy.
As each individual \captcha was served to three separate participants, we measured agreement between any two or more participants.
We also observed that agreement increases dramatically (20\% on average) if responses are treated as case insensitive, as shown in Table~\ref{tab:accuracy}.

\begin{table}[h!]
\centering
\footnotesize
\caption{Humans vs.\ bot solving time (seconds) and accuracy (percentage) for different \captcha types.}
\label{tab:bots_vs_humans}
\begin{tabularx}{\columnwidth}{l l l l l}
\toprule
      &       \multicolumn{2}{c}{\textbf{Human}}                    & \multicolumn{2}{c}{\textbf{Bot}}  \\\cmidrule(lr){2-3}\cmidrule(lr){4-5}
\textbf{\Captcha Type}      &  Time                  &   Accuracy       &  Time                  &  Accuracy        \\ \midrule
reCAPTCHA (click) &   3.1-4.9                             &  71-85\%               & 1.4 \cite{Sivakorn2016}      & 100\% \cite{Sivakorn2016}              \\ \midrule
Geetest           &  28-30                            &  N/A                   & 5.3~\cite{Haiqin2019}            & 96\%  ~\cite{Haiqin2019}               \\ \midrule
Arkose            &  18-42                            &  N/A                   & N/A                              & N/A                                    \\ \midrule
Distorted Text    &  9-15.3                            &  50-84\%               & \textless{}1 ~\cite{Zi20}       & 99.8\% \cite{Goodfellow}              \\ \midrule
reCAPTCHA (image) &  15-26                              &  81\%                  & 17.5  \cite{Hossen2020}        & 85\% \cite{Hossen2020}                 \\ \midrule
hCAPTCHA          &  18-32                            &  71-81\%               & 14.9 \cite{Hossen2021}           & 98\% \cite{Hossen2021}                 \\ \bottomrule
\end{tabularx}

\vspace{\floatsep}
\caption{Agreement for distorted text \captchas.}
\label{tab:accuracy}
\begin{tabularx}{\columnwidth}{X c c}
\toprule
      & Average Agreement & Average Agreement (case insensitive) \\
	   \midrule
Simple & 84\%             & 93\%                                                                      \\
Masked & 50\%             & 73\%                                                                      \\
Moving & 62\%             & 90\%                                                                      \\
\midrule
Total  & 65\%             & 85\%                                                                      \\
\bottomrule
\end{tabularx}
\end{table}

\section{Measuring User Abandonment}
\label{sec:eval_abandonment}
This subsection addresses \textbf{RQ5:} \emph{Does experimental context influence abandonment?}
Upon completion, we observed that the number of \captchas solved during our study exceeded what would be expected based on the number of participants who completed the study.
We hypothesized that this was due to participants starting but not completing the study.
To measure this behavior, we conducted a second user study that collected timestamps between \captchas, regardless of whether the entire study was completed.
We measured: (1) how many participants started the task; (2) how many abandoned the task when solving a \captcha; and (3) if so, at which task and \captcha.

This abandonment-focused study consisted of four groups, each with $100$ unique participants.
Two groups were presented with the  direct setting and the other two with the contextualized setting (see Section~\ref{sec:B_vs_UB_SD}).
We hypothesized that the amount of compensation might also impact abandonment, so we doubled the compensation for one of the groups in each setting.
The studies were run sequentially to avoid prospective participants simply picking the higher-paying study.

We summarize the key findings below, and present the full results in Tables~\ref{tab:unbiased_75},~\ref{tab:unbiased_150},~\ref{tab:biased_30}, and~\ref{tab:biased_60} in Appendix~\ref{sec:appendix-abandonment}.
Out of a total of $574$ participants who started the study, $174$ abandoned prior to completion (i.e., $30\%$ abandonment rate).
Several observations can be made:
First, in the direct setting, $25\%$ of the participants who ultimately abandoned the study did so before solving the first \captcha, but this rose to nearly $50\%$ in the contextualized setting.
Second, doubling the pay halved the abandonment rate for the contextualized setting (as expected), but increased it by $50\%$ in the direct setting.
Third, participants in the contextualized setting were $120\%$ more likely to abandon than those in the direct setting.
Fourth, in the contextualized setting, participants at the higher compensation level solved \captchas faster than those at the lower compensation level ($21.5\%$ decrease in average solving time across all \captcha types).
Interestingly, in the direct setting, participants at the higher compensation level solved \captchas \emph{slower} than those at the lower compensation level ($27.4\%$ \emph{increase} in average solving time across all \captcha types).
Finally, some \captcha types (e.g., Geetest) exhibited higher rates of abandonment than others.

This initial investigation strongly motivates the need for further exploration of \captcha-induced abandonment.
Although we studied the impact of compensation and experimental context, there may be other reasons behind abandonment, such as: \captcha type, \captcha difficulty, and expected duration of study.
Nevertheless, the trend of average users' unwillingness to solve a \captcha during account creation (even for monetary compensation) is a relevant finding for websites that choose to protect account creation (and/or account access) using \captchas.

%% file: content/06-related_work.tex
\section{Related Work}
\label{sec:related}

\Captchas are a well-studied topic, with several prior studies investigating both existing and novel \captcha schemes.

\subsection{Comparison of methodologies}
Table~\ref{tab:methodology} summarizes the key methodological aspects of prior \captcha user studies, from which the following observations can be made:

\begin{table*}[ht]
\footnotesize
\caption{Methodology and details of previous \captcha-related user studies.}
\label{tab:methodology}
\begin{tabularx}{\textwidth}{l p{24mm} p{13mm} p{22mm} X p{20mm} p{25mm}}
\toprule
                          & \textbf{\Captcha types}         & \textbf{Delivery medium} & \textbf{Measurements}                           & \textbf{Survey methods}           & \textbf{\Captcha source}      & \textbf{Compensation} (USD per \# \captchas) \\ \midrule
Ours                      & Text, Image, Game, Slider, Behavior   & MTurk              & Time, Agreement, Accuracy, Abandonment, Context & Demographics, Preference          & Alexa                             & \$0.30-\$1.50 per 10                     \\ \midrule
\cite{Bursztein}          & Text, Audio                           & MTurk, Website     & Time, Agreement                                 & Demographics                      & Alexa                             & \$0.02-\$0.50 per 24-39                  \\ \midrule
\cite{manarDynamic2014}   & DCG Captcha                           & MTurk              & Time, Accuracy                                  & Demographics, SUS                 & Newly proposed                    & \$0.50 per 4                             \\ \midrule
\cite{mohitAutomatic2019} & reCAPGen Audio                        & MTurk              & Time, Accuracy                                  & Demographics, Rating/Preference   & Newly proposed                    & \$4.00 per 60                            \\ \midrule
\cite{gaoEmerging2019}    &  3D/2D Text                           & MTurk              & Time, Accuracy                                  & Demographics, SUS                 & \cite{EICAP1,EICAP2,EICAP3}       & \$1.00 per 30                            \\ \midrule
\cite{Ho}                 & Text                                  & MTurk, DevilTyper  & Time, Accuracy, Abandonment                     & None                              & Major websites                    & \$0.03 per 15 (MTurk), 30.00 per 1.4 mil \\ \midrule
\cite{Bigham}             & Text, Audio, Interface                & Website            & Time, Accuracy                                  & Demographics, Preference          & Alexa                             & None                                     \\ \midrule
\cite{fidas2011}          & Text                                  & Website            & None                                            & Demographics, Rating/Preference   & Newly proposed                    & None                                     \\ \midrule
\cite{Gao}                & Jigsaw puzzle                         & Website            & Time, Accuracy                                  & Demographics, Preference          & Newly proposed                    & None                                     \\ \midrule
\cite{Krol}               & Text, Game, NoBot                     & Website            & Time                                            & Workload, Perceptions, Preference & None                              & None                                     \\ \midrule
\cite{senCAP}             & SenCAPTCHA, Text, Image, Audio, Video & MTurk              & Time                                            & Demographics, Preference, SUS     & Newly proposed, \cite{securimage} & \$1.25 per 9-15                          \\ \midrule
\cite{tanth19}            & Text, Behavior, Invisible, Game, Math & Unknown            & Time                                            & Demographics, Preference          & None                              & None                                     \\ \midrule
\cite{Ross}               & Sketcha                               & MTurk              & Time, Accuracy                                  & Demographics                      & Newly proposed                    & \$0.05-\$0.30 per 10-12                  \\ \bottomrule
\end{tabularx}
\end{table*}

\begin{itemize}[itemsep=.2mm] 
\item Most prior research has focussed on distorted text and newly-proposed \captcha schemes.

\item MTurk and proprietary websites have been the norm across \captcha user studies (except DevilTyper~\cite{Ho}).

\item Whilst almost all studies measured solving time, there is a bifurcation in terms of accuracy measurements: studies evaluating their own \captcha schemes or reimplementing existing schemes typically have direct access to accuracy results, whereas those evaluating unmodified deployed \captchas can only measure quantities such as agreement.

\item Most studies measured demographics and ratings or preferences.
Some studies also measured workload, open response (perceptions), and perceived usability.

\end{itemize}

\subsection{Detailed comparisons}
\label{sec:detailed_comparison}
We present detailed comparisons of our methodology and results with three representative prior \captcha studies.

\taggedpara{Bursztein et al.}~\cite{Bursztein} presented the first large-scale study on human \captcha solving performance.
Focussing on distorted text and audio \captchas, they used both MTurk and an underground \captcha-solving service to measure solving time and accuracy.
In terms of solving times, they found that it took on average $9.8$ and $28.4$ seconds to solve distorted text and audio \captchas respectively.
Although we did not evaluate audio \captchas (as we did not observe these in our website inspection), our results for distorted text \captchas broadly agree at $12.5$ on average.
Similarly to our study, they used \emph{agreement} between participants as a proxy for accuracy.
For distorted text, they observed $71\%$ agreement, which is in line with our observation of $75\%$ when averaging case sensitive and insensitive versions (see Table~\ref{tab:accuracy}).

\taggedpara{Feng et al.}~\cite{senCAP} presented senCAPTCHA, a new \captcha type using orientation sensors designed specifically for mobile devices with small screens.
They evaluated its security against brute-force and ML-based attacks, and its usability through two usability studies totalling 472 participants.
The second user study compared senCAPTCHA against text-, audio-, image-, and video-based \captchas, some of which were reimplemented for the study.
senCAPTCHA had the lowest median solving time ($5.02$ seconds), followed by image ($9.6$), video ($10.08$), text ($11.93$), and audio ($47.07$).
With the exception of click-based reCAPTCHA, it can be extrapolated that senCAPTCHA would have a lower solving time than the other \captcha types in our study.
In terms of preferences, most participants in their study preferred senCAPTCHA.
Out of the \captcha types in our study, senCAPTCHA most closely resembles the game-based \captchas, which supports our finding that game-based \captchas are generally preferred over text and image-based \captchas (see Figure~\ref{fig:pref_res}).

\taggedpara{Tanthavech and Nimkoompai}~\cite{tanth19} performed a 40-participant user study, measuring solving time for five \captcha types: click-based reCAPTCHA, text-, game-, math-based, and a newly-proposed invisible \captcha, which is essentially a honeypot for bots.
In terms of solving times, their distorted text measurement ($12$ seconds) is in the middle of our observed range ($9$-$15$ seconds), which is expected since it closely resembles our \emph{masked} type of distorted text.
Similarly, their click-based reCAPTCHA measurement ($3.1$ seconds) is on the boundary of our range ($3.1$-$4.9$), which suggests they may have configured the ``easier for users'' setting.
Their game-based \captcha appears to have a lower solving time than ours, but this is likely due to the type of game.
We did not observe or evaluate any math-based \captchas.
They also asked participants several post-study questions about the five \captcha types.
Interestingly, their participants ``enjoyed'' the game-based \captcha more than reCAPTCHA (click), which is the inverse of our findings (see Figure~\ref{fig:pref_res}), but may again be due to the different types of game.

\taggedpara{Overall}, where our study measured similar quantities to prior work, our findings broadly agree.
However, there is still a high degree of diversity in the sets of quantities measured in each study (e.g., types of \captchas, effect of experimental context), suggesting that a plurality of studies are needed to understand the full \captcha landscape.

\begin{table*}[!ht]
\footnotesize
\caption{Comparison of results from prior user studies evaluating \captchas: audio (A), behavior (B), distorted text (DT), game (G), honeypot (HP), image (I), math (M), service (S), slider (SL), video (V) and newly-proposed (New). Some studies used non-unique (NU) participants or MTurk (MT). * denotes reimplemented \captcha types.}
\label{tab:related_results}
\begin{tabularx}{\textwidth}{l X X l l}
\toprule
  & \textbf{Unique users} & \textbf{\Captchas solved} & \textbf{Average solving time (seconds)} & \textbf{Average accuracy} \\ \midrule
Ours                & 1,400 (MT)           & 14,000          & 9-15 (DT), 15-32 (I), 18-42 (G), 29 (SL), 3.1-4.9 (B) & 50-84\% (DT), 71-81\% (I), 71-85\% (B) \\ \midrule
\cite{Bursztein}          & 1,100-11,800 (MT)   & 318,000        & 9.8 (DT), 28.4 (A), 22.4 (S) & 71\% (DT), 31\% (A), 93\% (ebay DT) \\ \midrule
\cite{manarDynamic2014}   & 120                 & 480            & 8.5-16 (New), 17-47 (Attacks) & 16-100\% (New) \\ \midrule
\cite{mohitAutomatic2019} & 79                  & 4,740          & 9.6 (New) & 78.2\% (New) \\ \midrule
\cite{gaoEmerging2019}    & 120                 & 3,600          & 10 (3D-DT), 6.2-6.7 (DT) & 84\% (3D-DT), 92-96\% (DT) \\ \midrule
\cite{Ho}                 & 5,000 (NU), 44 (MT) & 1.4 mil, 7,500 & 8.5-12 (DT) & 79\%-89\%  (DT) \\ \midrule
\cite{Bigham}             & 162, 14 (Interface) & 2,350          & 9.9 (DT), 50.9 (Blind DT), 22.8 (A) & 80\% (DT), 39-43\% (A) \\ \midrule
\cite{fidas2011}          & 210                 & 210            & None & None \\ \midrule
\cite{Gao}                & 100                 & 300            & 4.9-6.4 (New) & 78\%-87.5\% (New) \\ \midrule
\cite{Krol}               & 87                  & 261            & 20 (DT), 29 (G), 70 (NoBot) & None \\ \midrule
\cite{senCAP}             & 436                 & 4,920          & 12 (DT), 47 (A), 9.6 (I*), 5 (New), 12 (V*) & None\\ \midrule
\cite{tanth19}            & 40                  & 200            & 12 (DT), 0 (HP), 3.1 (B), 8.2 (G \cite{yu2015automatic}), 4.1 (M \cite{HERNANDEZCASTRO2010141}) & None \\ \midrule
\cite{Ross}               & 558 (NU)            & 14,302         & 35 (New) & 42\%-88\% (New) \\ \bottomrule
\end{tabularx}
\end{table*}

\subsection{Summarized comparisons}
\label{sec:comparison_of_results}
In addition, Table~\ref{tab:related_results} presents a summarized comparison of our results with those of other prior studies.

\taggedpara{Solving Time:}
Overall, the average solving time in our study ranged from $3.6$ to $42.7$~seconds per \captcha, which is a larger range than that observed by Bursztein et al.~\cite{Bursztein} in 2010 ($9.8$ -- $28.4$~seconds) but is similar to the 2019 study by Feng et al.~\cite{senCAP} (medians ranging from $5.0$ to $47.1$~seconds).
Although direct comparison of solving times is not always meaningful, even for the same \captcha type (e.g., due to differing implementations or difficulty settings), we can identify a few trends.
Firstly, our measured solving times for the three types of distorted text \captchas ($9$-$15$ seconds\footnote{Unless otherwise stated, measurements refer to average solving time.}) are within the range of observations from prior studies ($6$-$20$ seconds).
We can therefore use this as a reference point for comparisons.
Secondly, with the exception of behavior-based \captchas, we observed that all other \captcha types took longer than distorted text.
Without considering newly-proposed \captcha types, this trend is consistent across most prior studies (with the exception of \cite{senCAP} and \cite{tanth19}).
Thirdly, although we do not evaluate any newly-proposed \captcha types, the times reported for these by other studies are typically faster than most of the \captcha types in our study, suggesting that there is scope for developing new \captcha types with lower solving times.
Finally, even in comparison to newly-proposed schemes, the behavior-based \captchas (e.g. reCAPTCHA click) appear to have the lowest solving times overall.

\taggedpara{Accuracy:}
For the case-sensitive setting, we observed a relatively broad range of accuracy (i.e., agreement) measurements for distorted text ($50$-$84\%$).
However, in the case-insensitive setting, our accuracy range narrows to $73$-$93$\%, which more closely aligns with prior studies, which have reported distorted text accuracies in the range $71$-$96$\%.
This suggests that both participants and prior studies have focussed on the case-insensitive setting.
In terms of deployed \captchas, \cite{Bursztein} reported an accuracy of $93\%$ for distorted text \captchas used by EBay in 2010.
This is higher than for the image-based \captchas we measured ($71$-$81\%$), suggesting that the latter may have increased in difficulty.

\taggedpara{Security:}
Table \ref{tab:bots_vs_humans} shows a comparison of our results to prior security analyses. %
Automated attacks on various \captcha schemes have been quite successful~\cite{Zi20,chen17,tang18,Gao16,li20,Goodfellow,unCaptcha,deepCRACK,Sano2013,Bursztein2011,Shekhar2021,Saumya2017,Darnstadt2014,Hossen2020,Hossen2021,Alqahtani2020,Haiqin2019,Lorenzi2012,Sivakorn2016}.
The bots' accuracy ranges from 85-100\%, with the majority above 96\%.
This substantially exceeds the human accuracy range we observed ($50$-$85\%$). %
Furthermore the bots' solving times are significantly lower in all cases, except reCAPTCHA (image), where human solving time ($18$ seconds) is similar to the bots' ($17.5$ seconds).
However, in the contextualized setting, human solving time rises to $22$ seconds, indicating that in this more natural setting, humans are slightly slower than bots.

%% file: content/07-conclusion.tex
\section{Summary \& Future Work}
\label{sec:conclusion}

\todo{This section could be removed and nothing would be lost. You would be better served using this space to fill our some of the other requested information.}

This paper explores currently-deployed \captchas via inspection of 200 popular websites and a series of user studies totalling $1,400$-participants.
For the research questions we posed at the outset, our results:
\begin{compactitem}
\item[\textbf{RQ1:}] show that there are significant differences in mean solving times between \captcha types.
\item[\textbf{RQ2:}] show that users' preference is not fully correlated with \captcha solving time.
\item[\textbf{RQ3:}] show that experimental context significantly influences \captcha solving times.
\item[\textbf{RQ4:}] confirm the previously-reported effects of age on solving time.
\item[\textbf{RQ5:}] confirm the high rates of abandonment due to \captcha-related tasks and identify that experimental context 
impacts abandonment.
\end{compactitem}

\noindent
We anticipate several directions for future work, including obtaining detailed measurements through a controlled user study, and further investigating the causes of abandonment.

%% file: content/09-acknowledgements.tex
\section{Acknowledgements}
We thank the anonymous reviewers for their valuable comments, and we are especially grateful to the shepherd for guiding us through several revisions.
The work of UCI authors was supported in part by: NSF Award \#:1840197, NSF Award \#:1956393, and NCAE-C CCR 2020 Award \#:H98230-20-1-0345. 
Yoshimichi Nakatsuka was supported in part by The Nakajima Foundation.
Andrew Paverd was supported in part by a US-UK Fulbright Cyber Security Scholar Award.

%% file: content/10-appendix.tex
\clearpage
\raggedbottom

\section{Abandonment measurement}
\label{sec:appendix-abandonment}
Tables~\ref{tab:unbiased_75},~\ref{tab:unbiased_150},~\ref{tab:biased_30}, and~\ref{tab:biased_60} show the results from four groups of participants from the secondary study which aimed to measure abandonment.
Columns represent the order of \captchas shown, while rows represent the \captcha type.
Cell values represent the number of MTurkers who abandoned.

\begin{table}[h]
	\centering
	\footnotesize
	\caption{\small{Abandonment in contextualized setting (\$0.75 payment)}}
	\label{tab:unbiased_75}
	\begin{tabularx}{\columnwidth}{lXXXXXXXXXXl}
	\toprule
	 & 1 & 2 & 3 & 4 & 5 & 6 & 7 & 8 & 9 & 10 & Total \\
	\midrule
	reCAPTCHA (easy) & 5 & 0 & 0 & 0 & 2 & 0 & 0 & 0 & 0 & 0 & 7 \\
	Geetest (slide) & 3 & 1 & 2 & 1 & 3 & 0 & 0 & 1 & 1 & 1 & 13 \\
	Arkose (selection) & 8 & 2 & 0 & 1 & 1 & 0 & 0 & 0 & 0 & 0 & 12 \\
	Arkose (rotation) & 2 & 1 & 1 & 0 & 1 & 1 & 0 & 0 & 0 & 0 & 6 \\
	Distorted text (simple) & 2 & 1 & 0 & 0 & 0 & 2 & 1 & 0 & 0 & 0 & 6 \\
	Distorted text (moving) & 0 & 1 & 2 & 1 & 1 & 0 & 1 & 0 & 1 & 0 & 7 \\
	reCAPTCHA (difficult) & 5 & 0 & 1 & 1 & 0 & 0 & 0 & 0 & 0 & 0 & 7 \\
	Distorted text (masked) & 4 & 2 & 1 & 0 & 0 & 0 & 0 & 0 & 0 & 0 & 7 \\
	hCAPTCHA (easy) & 2 & 2 & 2 & 0 & 1 & 0 & 0 & 0 & 0 & 0 & 7 \\
	hCAPTCHA (difficult) & 4 & 1 & 2 & 1 & 0 & 0 & 1 & 0 & 0 & 0 & 9 \\
	\midrule
	Total & 35 & 11 & 11 & 5 & 9 & 3 & 3 & 1 & 2 & 1 & 81 \\
	\bottomrule
	\end{tabularx}
\end{table}
\begin{table}[h]
	\centering
	\footnotesize
	\caption{\small{Abandonment in contextualized setting (\$1.50 payment)}}
	\label{tab:unbiased_150}
	\begin{tabularx}{\columnwidth}{lXXXXXXXXXXl}
	\toprule
	 & 1 & 2 & 3 & 4 & 5 & 6 & 7 & 8 & 9 & 10 & Total \\
	\midrule
	reCAPTCHA (easy) & 2 & 1 & 0 & 0 & 0 & 0 & 0 & 0 & 0 & 0 & 3 \\
	Geetest (slide) & 4 & 0 & 0 & 0 & 0 & 1 & 0 & 1 & 2 & 0 & 8 \\
	Arkose (selection) & 1 & 2 & 0 & 0 & 0 & 1 & 0 & 0 & 0 & 0 & 4 \\
	Arkose (rotation) & 4 & 0 & 1 & 0 & 0 & 0 & 0 & 1 & 0 & 0 & 6 \\
	Distorted text (simple) & 2 & 0 & 0 & 1 & 0 & 0 & 0 & 0 & 0 & 0 & 3 \\
	Distorted text (moving) & 1 & 1 & 1 & 0 & 0 & 1 & 1 & 0 & 0 & 0 & 5 \\
	reCAPTCHA (difficult) & 2 & 1 & 0 & 0 & 0 & 0 & 0 & 0 & 0 & 0 & 3 \\
	Distorted text (masked) & 1 & 2 & 0 & 0 & 0 & 0 & 0 & 0 & 0 & 0 & 3 \\
	hCAPTCHA (easy) & 1 & 1 & 0 & 0 & 0 & 0 & 0 & 0 & 0 & 0 & 2 \\
	hCAPTCHA (difficult) & 0 & 0 & 1 & 0 & 0 & 0 & 0 & 0 & 1 & 0 & 2 \\
	\midrule
	Total & 18 & 8 & 3 & 1 & 0 & 3 & 1 & 2 & 3 & 0 & 39 \\
	\bottomrule
	\end{tabularx}
\end{table}
\begin{table}[h]
	\centering
	\footnotesize
	\caption{\small{Abandonment in direct setting (\$0.30 payment)}}
	\label{tab:biased_30}
	\begin{tabularx}{\columnwidth}{lXXXXXXXXXXl}
	\toprule
	 & 1 & 2 & 3 & 4 & 5 & 6 & 7 & 8 & 9 & 10 & Total \\
	\midrule
	reCAPTCHA (easy) & 0 & 0 & 0 & 1 & 0 & 0 & 0 & 0 & 0 & 0 & 1 \\
	Geetest (slide) & 1 & 1 & 0 & 0 & 1 & 0 & 1 & 2 & 0 & 0 & 6 \\
	Arkose (selection) & 2 & 1 & 1 & 0 & 0 & 1 & 0 & 0 & 0 & 0 & 5 \\
	Arkose (rotation) & 0 & 0 & 0 & 0 & 0 & 1 & 0 & 0 & 0 & 0 & 1 \\
	Distorted text (simple) & 0 & 0 & 0 & 0 & 0 & 0 & 0 & 0 & 0 & 0 & 0 \\
	Distorted text (moving) & 0 & 0 & 0 & 0 & 0 & 1 & 1 & 0 & 0 & 0 & 2 \\
	reCAPTCHA (difficult) & 0 & 0 & 0 & 1 & 0 & 0 & 0 & 0 & 0 & 0 & 1 \\
	Distorted text (masked) & 0 & 0 & 0 & 0 & 0 & 0 & 1 & 0 & 0 & 0 & 1 \\
	hCAPTCHA (easy) & 1 & 1 & 0 & 1 & 0 & 0 & 0 & 0 & 0 & 0 & 3 \\
	hCAPTCHA (difficult) & 1 & 0 & 0 & 1 & 0 & 0 & 0 & 0 & 0 & 0 & 2 \\
	\midrule
	Total & 5 & 3 & 1 & 4 & 1 & 3 & 3 & 2 & 0 & 0 & 22 \\
	\bottomrule
	\end{tabularx}
\end{table}
\begin{table}[t]
	\centering
	\footnotesize
	\caption{\small{Abandonment in direct setting (\$0.60 payment)}}
	\label{tab:biased_60}
	\begin{tabularx}{\columnwidth}{lXXXXXXXXXXl}
	\toprule
	 & 1 & 2 & 3 & 4 & 5 & 6 & 7 & 8 & 9 & 10 & Total \\
	\midrule
	reCAPTCHA (easy) & 0 & 0 & 0 & 0 & 0 & 0 & 0 & 0 & 0 & 0 & 0 \\
	Geetest (slide) & 4 & 3 & 2 & 0 & 3 & 5 & 0 & 0 & 2 & 0 & 19 \\
	Arkose (selection) & 0 & 0 & 1 & 0 & 0 & 0 & 0 & 0 & 0 & 0 & 1 \\
	Arkose (rotation) & 1 & 0 & 0 & 2 & 1 & 0 & 0 & 0 & 0 & 0 & 4 \\
	Distorted text (simple) & 0 & 0 & 0 & 0 & 0 & 0 & 0 & 0 & 0 & 0 & 0 \\
	Distorted text (moving) & 1 & 0 & 0 & 0 & 0 & 0 & 0 & 0 & 1 & 0 & 2 \\
	reCAPTCHA (difficult) & 0 & 0 & 0 & 0 & 0 & 0 & 0 & 0 & 1 & 0 & 1 \\
	Distorted text (masked) & 2 & 0 & 0 & 0 & 0 & 0 & 0 & 0 & 0 & 0 & 2 \\
	hCAPTCHA (easy) & 0 & 1 & 0 & 1 & 0 & 0 & 0 & 0 & 0 & 0 & 2 \\
	hCAPTCHA (difficult) & 0 & 0 & 0 & 0 & 1 & 0 & 0 & 0 & 0 & 0 & 1 \\
	\midrule
	Total & 8 & 4 & 3 & 3 & 5 & 5 & 0 & 0 & 4 & 0 & 32 \\
	\bottomrule
	\end{tabularx}
\end{table}

\section{Questions asked in User Study}
\label{sec:appendix-questions}
Table~\ref{tab:questions} shows the exact questions that were asked to the participants during the pre- and post-study questionnaire.

\begin{table}[h!]
\footnotesize
\centering
\caption{Questions in user study}
\label{tab:questions}
\begin{tabularx}{\columnwidth}{p{0.58\linewidth} X}
\toprule
\textbf{Question} & \textbf{Possible Answers} \\ 
\midrule
\multicolumn{2}{c}{\emph{Pre-study questions}} \\
\midrule
Age & 18 - 100\\
\midrule
Gender & Male, Female, Non-binary\\
\midrule
What is your country of residence? & \emph{[selected from list of countries]}\\
\midrule
What is your highest level of Education? & No formal education, High School, Associate, Bachelor's, Master's, Doctorate\\
\midrule
Which of the following most closely describes the majority of your Internet use? & Work, Education, Browsing the Web, Gaming, Other \\
\midrule
Which device type are you using for this survey? & Phone, Computer (Desktop / Laptop), Tablet \\
\midrule
Which input method are you using for this survey? & Touchscreen, Keyboard, Other\\
\midrule
\textbf{\emph{[Only in the direct setting:]}}
Are you familiar with the purpose of CAPTCHAs? & Yes, No \\
\midrule
\multicolumn{2}{c}{\emph{Post-study question}} \\
\midrule
On a scale of 1-5, how enjoyable was solving the following CAPTCHA types? 
(1 being the least, and 5 -- the most, enjoyable). If the CAPTCHA type wasn't shown to you 
please put a 0 in that place. Note: You may not have seen the exact images shown, they are templates 
designed to represent different CAPTCHA types. & \emph{[single digit]} \\
\bottomrule
\end{tabularx}
\end{table}

\pagebreak

\section{Statistical Analysis of Solving Times}
\label{sec:eval_statistical}
To confirm the validity of our conclusions, we conducted several standard tests on the measured solving times.
We used the Holm-Bonferroni method to adjust for family-wise error in our statistical tests.
\begin{itemize}[itemsep=.2mm] 
\item First, we performed the \emph{Shapiro-Wilk normality test} with a null hypothesis that solving times adhere to a normal distribution.
For all \captcha types, results showed that we can reject the null hypothesis ($p < 0.001$).
\item Second, we ran a \emph{skewness test} with a null hypothesis that the skewness of the sample population is the same as that of a corresponding normal distribution.
For all \captcha types, results allowed us to reject the null hypothesis in favor of the alternative: the distribution of solving times is skewed ($p < 0.001$).
\item Third, we used the \emph{tailedness test} with a null hypothesis that the kurtosis of the sample population is the same as that of a normal distribution.
Results showed that, for all except distorted text (moving), the samples were drawn from a population that has a heavy-tailed distribution ($p < 0.001$).
\end{itemize}
Since solving times are: (1) not normally distributed, and (2) heavy tailed, we selected the \emph{Brown Forsythe test} to compare the equality of variance between different types of \captchas.
Results show that these distributions do not have equal variance, thus confirming our observations in Section~\ref{sec:eval_solving_time}.
Given the result of the Brown Forsythe test, we selected the \emph{Kruskal-Wallis test} to test the equality of mean.
For two pairs: reCAPTCHA (easy image) - hCAPTCHA (easy) and reCAPTCHA (easy click) - (hard click), we didn't see any statistical evidence that the means differ.
For the remainder, this test showed strong statistical evidence that the means differ ($p < 0.05$ between masked and moving distorted text and $p < 0.001$ for all other combinations). 

\pagebreak
\section{\Captcha Solving Times for Other Demographic Features}
\label{sec:dem_appendix}
\begin{figure}[h!]
        \centering
        \includegraphics[width=\columnwidth,trim=0mm 2mm 0 2mm,clip]{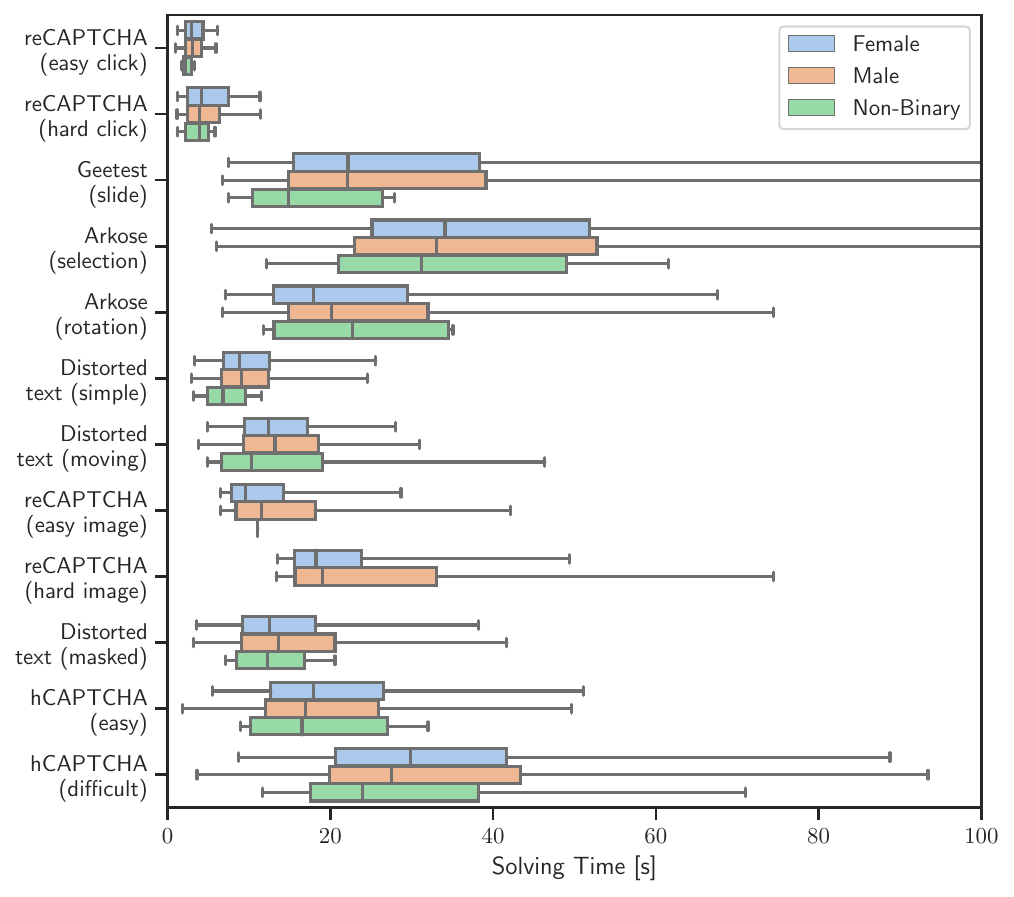}
        \caption{Effects of Gender.}
        \label{fig:gender_vs_time}
	\end{figure}

\begin{figure}[h!]
        \centering
        \includegraphics[width=\columnwidth,trim=0 2mm 0 2mm,clip]{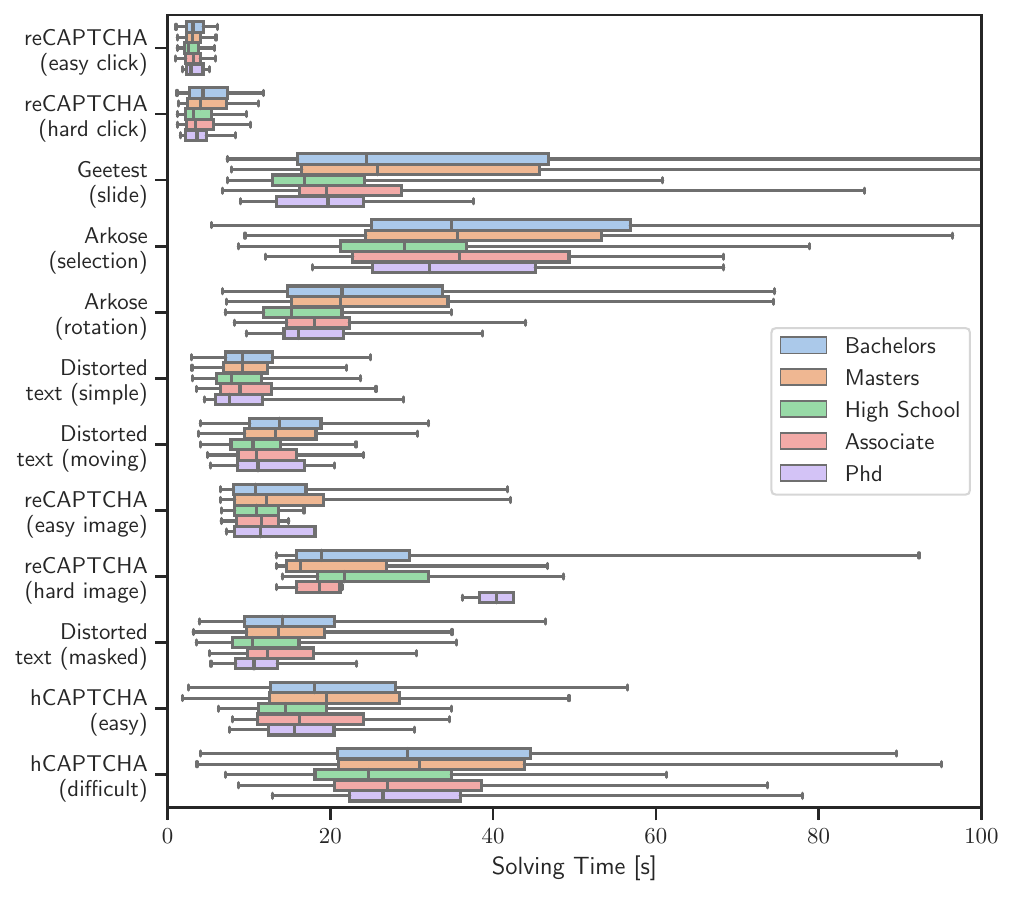}
        \caption{Effects of Education Level.}
        \label{fig:education_vs_time}
	\end{figure}

Figures~\ref{fig:gender_vs_time} and \ref{fig:education_vs_time} show participants' solving times analyzed across other demographic features.

%% file: CaptchasITW.bbl
\begin{thebibliography}{10}

\bibitem{360}
360.cn.
\newblock \url{{https://passport.360.cn/}}.

\bibitem{alexa}
{Alexa Top Sites}.
\newblock \url{{https://www.alexa.com/topsites}}.

\bibitem{mturk}
{Amazon Mechanical Turk}.
\newblock \url{{https://www.mturk.com/}}.

\bibitem{arkose}
{Arkose Labs}.
\newblock \url{{https://www.arkoselabs.com/about-us/}}.

\bibitem{cap_usage}
{CAPTCHA Usage Distribution on the Entire Internet}.
\newblock
  \url{{https://trends.builtwith.com/widgets/captcha/traffic/Entire-Internet}}.

\bibitem{ciscoUmbrella}
{Cisco Umbrella 1 Million}.
\newblock \url{{https://umbrella.cisco.com/blog/cisco-umbrella-1-million}}.

\bibitem{cloudflare_radar}
{Cloudflare Radar - Domain Rankings}.
\newblock \url{{https://radar.cloudflare.com/domains}}.

\bibitem{geetest}
{GeeTest CAPTCHA}.
\newblock \url{{https://www.geetest.com/en/Captcha}}.

\bibitem{hCaptcha}
{hCaptcha}.
\newblock \url{{https://www.hcaptcha.com/}}.

\bibitem{hCaptcha_largest}
{hCaptcha Is Now The Largest Independent CAPTCHA Service, Runs on 15\% Of The
  Internet}.
\newblock
  \url{{https://www.hcaptcha.com/post/hcaptcha-now-the-largest-independent-captcha-service}}.

\bibitem{reCAPTCHAv2}
{Invisible reCAPTCHA}.
\newblock \url{{https://developers.google.com/recaptcha/docs/invisible}}.

\bibitem{jrj}
jrj.com.
\newblock \url{{https://sso.jrj.com/}}.

\bibitem{nucaptcha}
{NuData Security}.
\newblock \url{{https://nudatasecurity.com/}}.

\bibitem{reCAPTCHA}
{reCAPTCHA}.
\newblock \url{{https://www.google.com/recaptcha/about/}}.

\bibitem{reCAPTCHAv3}
{reCAPTCHA v3}.
\newblock \url{{https://developers.google.com/recaptcha/docs/v3}}.

\bibitem{majestic}
{The Majestic Million}.
\newblock \url{{https://majestic.com/reports/majestic-million}}.

\bibitem{tor}
{The Tor Project: Privacy \& Freedom Online}.
\newblock \url{{https://www.torproject.org/}}.

\bibitem{Xinhuanet}
Xinhuanet.
\newblock \url{{https://mail.xinhuanet.com}}.

\bibitem{mongodb}
{MongoDB}.
\newblock \url{{https://www.mongodb.com/}}, 2021.

\bibitem{nodejs}
{Node.js}.
\newblock \url{{https://nodejs.org/}}, 2021.

\bibitem{deepCRACK}
W.~Aiken and H.~Kim.
\newblock {POSTER: DeepCRACk: Using Deep Learning to Automatically CRack Audio
  CAPTCHAs}.
\newblock In {\em Proceedings of the 2018 on Asia Conference on Computer and
  Communications Security}, ASIACCS '18, page 797–799, New York, NY, USA,
  2018. ACM.

\bibitem{Alqahtani2020}
F.~H. Alqahtani and F.~A. Alsulaiman.
\newblock {Is image-based CAPTCHA secure against attacks based on machine
  learning? An experimental study}.
\newblock {\em Computers \& Security}, 88:101635, 2020.

\bibitem{belk}
M.~Belk, P.~Germanakos, C.~Fidas, A.~Holzinger, and G.~Samaras.
\newblock {Towards the Personalization of CAPTCHA Mechanisms Based on
  Individual Differences in Cognitive Processing}.
\newblock In A.~Holzinger, M.~Ziefle, M.~Hitz, and M.~Debevc, editors, {\em
  Human Factors in Computing and Informatics}, pages 409--426, Berlin,
  Heidelberg, 2013. Springer Berlin Heidelberg.

\bibitem{Bigham}
J.~P. Bigham and A.~Cavender.
\newblock {Evaluating Existing Audio CAPTCHAs and an Interface Optimized for
  Non-Visual Use}.
\newblock In {\em Proceedings of the SIGCHI Conference on Human Factors in
  Computing Systems}, CHI '09, page 1829–1838, New York, NY, USA, 2009. ACM.

\bibitem{unCaptcha}
K.~Bock, D.~Patel, G.~Hughey, and D.~Levin.
\newblock {unCaptcha: A Low-Resource Defeat of reCaptcha{\textquoteright}s
  Audio Challenge}.
\newblock In {\em 11th {USENIX} Workshop on Offensive Technologies ({WOOT}
  17)}, Vancouver, BC, Aug. 2017. {USENIX} Association.

\bibitem{Bursztein2011}
E.~Bursztein, R.~Beauxis, H.~Paskov, D.~Perito, C.~Fabry, and J.~Mitchell.
\newblock {The Failure of Noise-Based Non-continuous Audio Captchas}.
\newblock In {\em 2011 IEEE Symposium on Security and Privacy}, pages 19--31,
  2011.

\bibitem{Bursztein}
E.~Bursztein, S.~Bethard, C.~Fabry, J.~C. Mitchell, and D.~Jurafsky.
\newblock {How Good Are Humans at Solving CAPTCHAs? A Large Scale Evaluation}.
\newblock In {\em IEEE Symposium on Security and Privacy}, 2010.

\bibitem{rfc1262}
V.~G. Cerf.
\newblock {Guidelines for Internet Measurement Activities}.
\newblock RFC 1262, Oct. 1991.

\bibitem{chen17}
J.~Chen, X.~Luo, Y.~Guo, Y.~Zhang, and D.~Gong.
\newblock {A Survey on Breaking Technique of Text-Based CAPTCHA}.
\newblock {\em Security and Communication Networks}, 12 2017.

\bibitem{chmielewski2020mturk}
M.~Chmielewski and S.~C. Kucker.
\newblock {An MTurk crisis? Shifts in data quality and the impact on study
  results}.
\newblock {\em Social Psychological and Personality Science}, 11(4):464--473,
  2020.

\bibitem{forrester2021}
F.~Consulting.
\newblock State of online fraud and bot management.
\newblock
  \url{https://services.google.com/fh/files/misc/google_forrester_bot_management_tlp_post_production_final.pdf},
  2021.

\bibitem{Darnstadt2014}
M.~Darnst{\"a}dt, H.~Meutzner, and D.~Kolossa.
\newblock {Reducing the Cost of Breaking Audio CAPTCHAs by Active and
  Semi-supervised Learning}.
\newblock In {\em 2014 13th International Conference on Machine Learning and
  Applications}, pages 67--73, 2014.

\bibitem{senCAP}
Y.~Feng, Q.~Cao, H.~Qi, and S.~Ruoti.
\newblock Sencaptcha: A mobile-first captcha using orientation sensors.
\newblock {\em Proc. ACM Interact. Mob. Wearable Ubiquitous Technol.}, 4(2),
  jun 2020.

\bibitem{fidas2011}
C.~A. Fidas, A.~G. Voyiatzis, and N.~M. Avouris.
\newblock {On the Necessity of User-Friendly CAPTCHA}.
\newblock In {\em Proceedings of the SIGCHI Conference on Human Factors in
  Computing Systems}, CHI '11, page 2623–2626, New York, NY, USA, 2011. ACM.

\bibitem{gao2012divide}
H.~Gao, W.~Wang, and Y.~Fan.
\newblock {Divide and conquer: an efficient attack on Yahoo! CAPTCHA}.
\newblock In {\em 2012 IEEE 11th International Conference on Trust, Security
  and Privacy in Computing and Communications}, pages 9--16. IEEE, 2012.

\bibitem{Gao16}
H.~Gao, J.~Yan, F.~Cao, Z.~Zhang, L.~Lei, M.~Tang, P.~Zhang, X.~Zhou, X.~Wang,
  and J.~Li.
\newblock {A Simple Generic Attack on Text Captchas}.
\newblock In {\em {Network and Distributed System Security Symposium (NDSS)}},
  San Diego, California, United States, 2016.

\bibitem{Gao}
H.~{Gao}, D.~{Yao}, H.~{Liu}, X.~{Liu}, and L.~{Wang}.
\newblock {A Novel Image Based CAPTCHA Using Jigsaw Puzzle}.
\newblock In {\em 2010 13th IEEE International Conference on Computational
  Science and Engineering}, pages 351--356, 2010.

\bibitem{gaoEmerging2019}
S.~Gao, M.~Mohamed, N.~Saxena, and C.~Zhang.
\newblock {Emerging-Image Motion CAPTCHAs: Vulnerabilities of Existing Designs,
  and Countermeasures}.
\newblock {\em IEEE Transactions on Dependable and Secure Computing},
  16(6):1040--1053, 2019.

\bibitem{Goodfellow}
I.~J. Goodfellow, Y.~Bulatov, J.~Ibarz, S.~Arnoud, and V.~Shet.
\newblock {Multi-digit number recognition from street view imagery using deep
  convolutional neural networks}.
\newblock {\em arXiv preprint arXiv:1312.6082}, 2014.

\bibitem{Guerar}
M.~Guerar, L.~Verderame, M.~Migliardi, F.~Palmieri, and A.~Merlo.
\newblock {Gotta {CAPTCHA} 'Em All: {A} Survey of Twenty years of the
  Human-or-Computer Dilemma}.
\newblock {\em CoRR}, abs/2103.01748, 2021.

\bibitem{hernandez2010pitfalls}
C.~J. Hernandez-Castro and A.~Ribagorda.
\newblock {Pitfalls in CAPTCHA design and implementation: The Math CAPTCHA, a
  case study}.
\newblock {\em Computers \& Security}, 29(1):141--157, 2010.

\bibitem{HERNANDEZCASTRO2010141}
C.~J. Hernandez-Castro and A.~Ribagorda.
\newblock Pitfalls in captcha design and implementation: The math captcha, a
  case study.
\newblock {\em Computers \& Security}, 29(1):141--157, 2010.

\bibitem{Ho}
C.-J. Ho, C.-C. Wu, K.-T. Chen, and C.-L. Lei.
\newblock {DevilTyper: A Game for CAPTCHA Usability Evaluation}.
\newblock {\em Comput. Entertain.}, 9(1), apr 2011.

\bibitem{Hossen2021}
M.~I. Hossen and X.~Hei.
\newblock {A Low-Cost Attack against the hCaptcha System}.
\newblock {\em CoRR}, abs/2104.04683, 2021.

\bibitem{Hossen2020}
M.~I. Hossen, Y.~Tu, M.~F. Rabby, M.~N. Islam, H.~Cao, and X.~Hei.
\newblock {An Object Detection based Solver for Google{\textquoteright}s Image
  reCAPTCHA v2}.
\newblock In {\em 23rd International Symposium on Research in Attacks,
  Intrusions and Defenses ({RAID} 2020)}, pages 269--284, San Sebastian, Oct.
  2020. {USENIX} Association.

\bibitem{imperva2022}
Imperva.
\newblock Imperva bad bot report.
\newblock
  \url{https://www.imperva.com/resources/resource-library/reports/bad-bot-report/},
  2022.

\bibitem{mohitAutomatic2019}
M.~Jain, R.~Tripathi, I.~Bhansali, and P.~Kumar.
\newblock {Automatic Generation and Evaluation of Usable and Secure Audio
  ReCAPTCHA}.
\newblock In {\em The 21st International ACM SIGACCESS Conference on Computers
  and Accessibility}, ASSETS '19, page 355–366, New York, NY, USA, 2019.
  Association for Computing Machinery.

\bibitem{Krol}
K.~Krol, S.~Parkin, and M.~A. Sasse.
\newblock {Better the Devil You Know: A User Study of Two CAPTCHAs and a
  Possible Replacement Technology}.
\newblock In {\em 2016 NDSS Workshop on Usable Security}, pages 1--10, 2016.

\bibitem{li20}
C.~Li, X.~Chen, H.~Wang, P.~Wang, Y.~Zhang, and W.~Wang.
\newblock {End-to-end attack on text-based CAPTCHAs based on cycle-consistent
  generative adversarial network}.
\newblock {\em Neurocomputing}, 433:223--236, 2021.

\bibitem{Lorenzi2012}
D.~Lorenzi, J.~Vaidya, E.~Uzun, S.~Sural, and V.~Atluri.
\newblock {Attacking Image Based CAPTCHAs Using Image Recognition Techniques}.
\newblock In V.~Venkatakrishnan and D.~Goswami, editors, {\em Information
  Systems Security}, pages 327--342, Berlin, Heidelberg, 2012. Springer Berlin
  Heidelberg.

\bibitem{EICAP1}
N.~Mitra, H.~Chu, T.~Lee, L.~Wolf, H.~Yeshurun, and D.~Cohen-Or.
\newblock Emerging images.
\newblock In {\em Proceedings of ACM SIGGRAPH Asia 2009, SIGGRAPH Asia '09},
  volume~28, pages 163:1--163:8, 2009.
\newblock ACM SIGGRAPH Asia 2009, SIGGRAPH Asia '09 ; Conference date:
  16-12-2009 Through 19-12-2009.

\bibitem{manarDynamic2014}
M.~Mohamed, S.~Gao, N.~Saxena, and C.~Zhang.
\newblock {Dynamic Cognitive Game CAPTCHA Usability and Detection of
  Streaming-Based Farming}.
\newblock In {\em 2014 NDSS Workshop on Usable Security}, pages 1--10, 2014.

\bibitem{moss_2020}
A.~Moss.
\newblock {After the bot scare: Understanding what's been happening with data
  collection on MTurk and how to stop it}.
\newblock
  \url{{https://www.cloudresearch.com/resources/blog/after-the-bot-scare-understanding-whats-been-happening-with-data-collection-on-mturk-and-how-to-stop-it/}},
  Aug 2020.

\bibitem{Motoyama}
M.~Motoyama, K.~Levchenko, C.~Kanich, D.~McCoy, G.~M. Voelker, and S.~Savage.
\newblock {Re: CAPTCHAs{\textemdash}Understanding CAPTCHA-Solving Services in
  an Economic Context}.
\newblock In {\em 19th {USENIX} Security Symposium ({USENIX} Security 10)},
  Washington, DC, aug 2010. {USENIX} Association.

\bibitem{securimage}
D.~Phillips.
\newblock Secureimage: {PHP CAPTCHA} script.
\newblock \url{https://www.phpcaptcha.org/}, 2023.

\bibitem{pochat2019tranco}
V.~L. Pochat, T.~van Goethem, S.~Tajalizadehkhoob, M.~Korczynski, and
  W.~Joosen.
\newblock Tranco: {A} research-oriented top sites ranking hardened against
  manipulation.
\newblock In {\em 26th Annual Network and Distributed System Security
  Symposium, {NDSS} 2019, San Diego, California, USA, February 24-27, 2019}.
  The Internet Society, 2019.

\bibitem{cloudflare}
M.~Prince and S.~Isasi.
\newblock {Moving from reCAPTCHA to hCaptcha}.
\newblock
  \url{{https://blog.cloudflare.com/moving-from-recaptcha-to-hcaptcha/}}.

\bibitem{Ross}
S.~A. Ross, J.~A. Halderman, and A.~Finkelstein.
\newblock {Sketcha: A Captcha Based on Line Drawings of 3D Models}.
\newblock In {\em Proceedings of the 19th International Conference on World
  Wide Web}, page 821–830, New York, NY, USA, 2010. ACM.

\bibitem{Sano2013}
S.~Sano, T.~Otsuka, and H.~G. Okuno.
\newblock {Solving Google's Continuous Audio CAPTCHA with HMM-Based Automatic
  Speech Recognition}.
\newblock In K.~Sakiyama and M.~Terada, editors, {\em Advances in Information
  and Computer Security}, pages 36--52, Berlin, Heidelberg, 2013. Springer
  Berlin Heidelberg.

\bibitem{Scheitle2018}
Q.~Scheitle, O.~Hohlfeld, J.~Gamba, J.~Jelten, T.~Zimmermann, S.~D. Strowes,
  and N.~Vallina-Rodriguez.
\newblock A long way to the top: Significance, structure, and stability of
  internet top lists.
\newblock In {\em Proceedings of the Internet Measurement Conference 2018}, IMC
  '18, page 478–493, New York, NY, USA, 2018. ACM.

\bibitem{Shekhar2021}
H.~Shekhar.
\newblock {Breaking Audio Captcha using Machine Learning/Deep Learning and
  Related Defense Mechanism}.
\newblock {\em San Jose State University Master's Projects}, 2019.

\bibitem{Shet}
V.~Shet.
\newblock {Street View and reCAPTCHA technology just got smarter}.
\newblock
  \url{{https://security.googleblog.com/2014/04/street-view-and-recaptcha-technology.html}},
  2014.

\bibitem{Sivakorn2016}
S.~Sivakorn, I.~Polakis, and A.~D. Keromytis.
\newblock {I am Robot: (Deep) Learning to Break Semantic Image CAPTCHAs}.
\newblock In {\em 2016 IEEE European Symposium on Security and Privacy (EuroS
  P)}, pages 388--403, 2016.

\bibitem{Saumya2017}
S.~Solanki, G.~Krishnan, V.~Sampath, and J.~Polakis.
\newblock {\em {In (Cyber)Space Bots Can Hear You Speak: Breaking Audio
  CAPTCHAs Using OTS Speech Recognition}}, page 69–80.
\newblock ACM, New York, NY, USA, 2017.

\bibitem{tang18}
M.~Tang, H.~Gao, Y.~Zhang, Y.~Liu, P.~Zhang, and P.~Wang.
\newblock {Research on Deep Learning Techniques in Breaking Text-Based Captchas
  and Designing Image-Based Captcha}.
\newblock {\em IEEE Transactions on Information Forensics and Security},
  13(10):2522--2537, 2018.

\bibitem{tanth19}
N.~Tanthavech and A.~Nimkoompai.
\newblock Captcha: Impact of website security on user experience.
\newblock {\em ICIIT '19: Proceedings of the 2019 4th International Conference
  on Intelligent Information Technology}, pages 37--41, 02 2019.

\bibitem{Uzun}
E.~Uzun, S.~Chung, I.~Essa, and W.~Lee.
\newblock {rtCaptcha: A Real-Time Captcha Based Liveness Detection System}.
\newblock In {\em {Network and Distributed System Security Symposium (NDSS)}},
  San Diego, California, United States, 02 2018.

\bibitem{vonAhn}
L.~von Ahn, M.~Blum, N.~J. Hopper, and J.~Langford.
\newblock {CAPTCHA: Using Hard AI Problems for Security}.
\newblock In E.~Biham, editor, {\em Advances in Cryptology --- EUROCRYPT 2003},
  pages 294--311, Berlin, Heidelberg, 2003. Springer Berlin Heidelberg.

\bibitem{webb2022too}
M.~A. Webb and J.~P. Tangney.
\newblock Too good to be true: Bots and bad data from mechanical turk.
\newblock {\em Perspectives on Psychological Science}, 2022.

\bibitem{Haiqin2019}
H.~Weng, B.~Zhao, S.~Ji, J.~Chen, T.~Wang, Q.~He, and R.~Beyah.
\newblock {Towards understanding the security of modern image captchas and
  underground captcha-solving services}.
\newblock {\em Big Data Mining and Analytics}, 2(2):118--144, 2019.

\bibitem{xie2022secrank}
Q.~Xie, S.~Tang, X.~Zheng, Q.~Lin, B.~Liu, H.~Duan, and F.~Li.
\newblock Building an open, robust, and stable voting-based domain top list.
\newblock In K.~R.~B. Butler and K.~Thomas, editors, {\em 31st {USENIX}
  Security Symposium, {USENIX} Security 2022, Boston, MA, USA, August 10-12,
  2022}, pages 625--642. {USENIX} Association, 2022.

\bibitem{EICAP3}
Y.~Xu, G.~Reynaga, S.~Chiasson, J.-M. Frahm, F.~Monrose, and P.~Van~Oorschot.
\newblock Security and usability challenges of moving-object captchas: Decoding
  codewords in motion.
\newblock In {\em Proceedings of the 21st USENIX Conference on Security
  Symposium}, Security'12, page~4, USA, 2012. USENIX Association.

\bibitem{EICAP2}
Y.~Xu, G.~Reynaga, S.~Chiasson, J.-M. Frahm, F.~Monrose, and P.~C. van
  Oorschot.
\newblock Security analysis and related usability of motion-based captchas:
  Decoding codewords in motion.
\newblock {\em IEEE Transactions on Dependable and Secure Computing},
  11(5):480--493, 2014.

\bibitem{yan2008low}
J.~Yan and A.~S. El~Ahmad.
\newblock {A Low-cost Attack on a Microsoft CAPTCHA}.
\newblock In {\em Proceedings of the 15th ACM conference on Computer and
  communications security}, pages 543--554, 2008.

\bibitem{yan2008}
J.~Yan and A.~S. El~Ahmad.
\newblock Usability of captchas or usability issues in captcha design.
\newblock In {\em Proceedings of the 4th Symposium on Usable Privacy and
  Security}, SOUPS '08, page 44–52, New York, NY, USA, 2008. ACM.

\bibitem{yu2015automatic}
H.~Yu and M.~O. Riedl.
\newblock Automatic generation of game-based captchas.
\newblock In {\em Proceedings of the FDG workshop on Procedural Content
  Generation}, 2015.

\bibitem{Zi20}
Y.~Zi, H.~Gao, Z.~Cheng, and Y.~Liu.
\newblock {An End-to-End Attack on Text CAPTCHAs}.
\newblock {\em IEEE Transactions on Information Forensics and Security},
  15:753--766, 2020.

\end{thebibliography}
